\documentclass[12pt,manuscript]{aastex}

\input psfig
\def\puncspace{\ifmmode\,\else{\ifcat.\C{\if.\C\else\if,\C\else\if?\C\else%
\if:\C\else\if;\C\else\if-\C\else\if)\C\else\if/\C\else\if]\C\else\if'\C%
\else\space\fi\fi\fi\fi\fi\fi\fi\fi\fi\fi}%
\else\if\empty\C\else\if\space\C\else\space\fi\fi\fi}\fi}
\def\SP{\let\\=\empty\futurelet\C\puncspace}
\def\etal{et\SP al. \SP }

\def\kms{km s$^{-1}$}
\def\h-1{$h^{-1}$}

\def\void#1{{}}

\def\h1{$h^{-1}$}

\def\etal{et al.\,}
\def\eg{e.g., \,}
\def\mb{$m_{B(0)}$\,}

\def\lsim{~\rlap{$<$}{\lower 1.0ex\hbox{$\sim$}}}
\def\gsim{~\rlap{$>$}{\lower 1.0ex\hbox{$\sim$}}}
\def\hmu2 {$\bar{\mu}_{1/2}$\SP}
\def\r2 {$r_{1/2}$\SP}
\def\mue {$<{\mu_e}>$\SP}
\def\dn {$D_n$\SP}
\def\ldn {$log\,D_n$\SP}
\def\re {$r_e$\SP}
\def\lre {$\log\,r_e$\SP}

\def\bri {mag~arcsec$^{-2}$\SP}

\def\m26 {$m_{26}$\SP}
\def\d26 {$D_{26}$\SP}

\def\mssrs2 {$m_{ssrs2}$\SP}

\def\AandA#1,{A\&A, {#1},}
\def\AAS#1,{A\&AS, {#1},}
\def\ARAA#1,{ARA\&A, {#1},}
\def\AJ#1,{AJ, {#1},}
\def\ApJ#1,{ApJ, {#1},}
\def\ApJL#1,{ApJ, {#1},}
\def\ApJS#1,{ApJS, {#1},}
\def\MN#1,{MNRAS, {#1},}
\def\MNRAS#1,{MNRAS, {#1},}
\def\PASP#1,{PASP, {#1},}
\def\Science#1,{Science, {#1},}
\def\Nature#1,{Nature, {#1},}
\def\ZfA#1,{{\sl Zeitschrift f\"ur Astrophysik}, {#1},}

\def\dnsig{$D_n-\sigma$\SP}

\shorttitle{Redshift-Distance Survey of Early-Type Galaxies}
\shortauthors{Alonso et al.}

\begin{document}

\title{Redshift-Distance Survey of Early-Type Galaxies: Circular 
Aperture Photometry\footnote{Based on observations at 
Cerro Tololo Interamerican Observatory (CTIO), operated by the
National Optical Astronomical Observatories, under AURA; European
Southern Observatory (ESO); Fred
Lawrence Whipple Observatory (FLWO); and
the MDM Observatory on Kitt Peak.}}
\author
{M. V. Alonso\altaffilmark{1,}\altaffilmark{2}, M. Bernardi\altaffilmark{3},
 L. N. da Costa\altaffilmark{4,}\altaffilmark{5}, G. Wegner,\altaffilmark{6}}
\author{C. N. A. Willmer\altaffilmark{5,}\altaffilmark{7}, 
P. S. Pellegrini\altaffilmark{5,}\altaffilmark{8} 
and M. A. G. Maia\altaffilmark{5,}\altaffilmark{8}}
\affil{\altaffiltext{1}{Observatorio Astr\'onomico de
C\'ordoba,  Laprida  854, C\'ordoba, 5000, Argentina and CONICET, Argentina}}
\affil{\altaffiltext{2}{CNRS UMR 5572, Observatoire Midi-Pyr\'en\'ees, 14 Avenue E. Belin, 31400 Toulouse, France}}
\affil{\altaffiltext{3}{Department of Physics, Carnegie Mellon University, 
5000 Forbes Ave., Pittsburgh, PA 15213, USA}}
\affil{\altaffiltext{4}{European Southern Observatory,
Karl-Schwarzschild Strasse 2, D-85748 Garching, Germany}}
\affil{\altaffiltext{5}{Observat\'orio Nacional, Rua General Jos\'e Cristino
77, Rio de Janeiro, R. J., 20921, Brazil}}
\affil{\altaffiltext{6}{Department of Physics \& Astronomy, Dartmouth
College, Hanover, NH  03755-3528, USA}}
\affil{\altaffiltext{7}{UCO/Lick Observatory, University of California,
1156 High Street, Santa Cruz,  CA 95064, USA}}
\affil{\altaffiltext{8}{Observatorio do Valongo, UFRJ, Ladeira do Pedro Antonio 43, Rio de Janeiro, R. J.,
20080-090, Brazil}}

\email{vicky@ast.obs-mip.fr (MVA)}
\date{Received \today}

\begin{abstract}
We present R-band CCD photometry for 1332 early-type galaxies,
observed as part of the ENEAR survey of peculiar motions using
early-type galaxies in the nearby Universe. Circular apertures are
used to trace the surface brightness profiles, which are then fit by a
two-component bulge-disk model. From the fits we obtain the structural
parameters required to estimate galaxy distances using the
$D_n-\sigma$ and Fundamental Plane relations.  We find that about 12
\% of the galaxies are well represented by a pure $r^{1/4}$ law while
87\% are best fit by a two component model.
There are 356 repeated observations of 257 galaxies obtained during
different runs that are used to derive statistical corrections and
bring the data to a common system. We also use these repeated
observations to estimate our internal errors. 
The accuracy of our measurements are tested by the comparison of
354 galaxies in common with other authors. Typical errors in our
measurements are 0.011 dex for $\log{D_n}$, 0.064 dex for $\log{r_e}$,
0.086~\bri for \mue and 0.09 for $m_{R_C}$, comparable to those estimated 
by other authors.
The photometric data reported here represent one of the largest
 high-quality and uniform all-sky samples currently available for
early-type galaxies in the nearby universe, especially suitable for
peculiar motion studies.

\end {abstract}


\section{Introduction}

With the aim of mapping
the distribution of total matter in the nearby Universe, we
have completed a redshift-distance survey of early-type
galaxies drawn from an all-sky magnitude-limited sample
(hereafter ENEAR, da Costa \etal 2000a), which is being used to map
the peculiar velocity field of galaxies in a volume of about 7000 \kms.
The current survey extends the earlier effort of Lynden-Bell \etal
(1988, hereafter 7S) by using a sample that is about three times larger,
reaches more than one magnitude fainter, and
also includes lenticular galaxies. In this paper we describe the
measurements and present the results
of the CCD photometry for the 1332 elliptical and lenticular
galaxies that were measured in circular apertures for the ENEAR survey.

The peculiar velocity field of galaxies is a
means of probing the total distribution of matter in the Universe
within the gravitational instability framework. While other methods are known,
in order to map the peculiar velocity field using galaxies, it is  
necessary to estimate distances that are 
independent of redshift; these distance determinations use scaling relations
between spectroscopically measurable distance independent 
properties and photometrically defined distance dependent
quantities such as radius or brightness.  
In addition, 
it is possible to compare independent determinations of the velocity field
using different galaxy samples and distance determination techniques,
which give strong support to the results.  For example, the well known 
Tully-Fisher relation (hereafter TF, Tully \& Fisher 1977, Mathewson 
\etal 1992, 1996, da Costa \etal 1996, 
Haynes \etal 1999a,b, Willick \etal 1997) has been extensively used
to map the velocity field using spiral galaxies.  Thus, employing 
early-type galaxies is a complementary analysis since a different observational
technique is used and early-type and spiral galaxies probe different regions 
of space. 

For early-type galaxies there are two scaling relations: (1) the Fundamental 
Plane (FP, Djorgovski \& Davis 1987), a three dimensional space defined by 
surface brightness (\mue), effective radius (\re), and central velocity
 dispersion ($\sigma$) and (2) the \dnsig relation (Dressler et al. 1987, 7S) 
which correlates the characteristic size (\dn) with $\sigma$. The \dnsig 
relation which can be shown to be a projection of the FP (J\o rgensen \etal 
1993) is accurate over a range of \mue and it is easier to apply in practice. 
We have used the \dnsig relation to estimate distances 
and derive the peculiar velocity field for the ENEAR survey because
\dn is simple to measure (it is obtained from an accurate interpolation, 
relying neither on fits to the galaxy light profiles nor on extrapolations)
and it is as accurate as using \re and \mue in the FP. Although, since the 
\dnsig relation is an approximation to the FP, it could introduce extra 
scatter in the measurements compared to using the FP. 
The resulting \dnsig relation
can then be easily constructed using galaxies in clusters.
The ENEAR template \dnsig relation,
obtained from the subsample of galaxies in clusters, has been discussed
in Bernardi et al. (2002a, 2002b), where the scatter in the present 
sample implies a distance error of about 19\% per galaxy.

The photometric data reported here uses the CCD imaging of the
ENEAR survey galaxies measured in circular apertures.
Galaxy parameters for both \dnsig and the FP
are determined using the Saglia et al. (1997)
two-component disk-bulge model, which accounts
for the smearing of light due to seeing and employs
a sequence of $r^{1/4}$
and exponential profiles with appropriate scale lengths and disk-to-bulge
ratios.  The excellent fits achieved here
using the two component model profile justify its use in determining the
FP parameters.

As part of the papers of the series, the central velocity dispersion and
line strengths of the galaxies obtained from the spectroscopic data of
the survey are presented in  Wegner et al. (2003). A number of results 
using the ENEAR data (photometric and spectroscopic data) have already 
appeared. These
include statistical analyses of the ENEAR sample using
the velocity correlation function (Borgani et al. 2000) and the
dipole measurements (da Costa et al. 2000b). These are in good
agreement with the results obtained by the TF surveys,
demonstrating that estimates of the
cosmic flow as traced by early-type and spiral galaxies are statistically
equivalent even though both galaxy types obey distinct distance relations
and sample different density regimes, but probe the same induced
peculiar velocity field. Nusser et al. (2001), examined  the peculiar 
velocity field using the ENEAR  galaxies and the PSCz gravity fields
and showed that
the likelihood analysis of the ENEAR and PSCz modes are in good agreement
with the values obtained from the TF surveys.
Further analyses include the measurement of the large-scale
power spectrum obtained from the ENEAR peculiar velocity
field (Zaroubi et al. 2001).  In general, all these results suggest
low-amplitude bulk flows and that most of the motion of the
Local Group is due to mass fluctuations within the volume of \nobreak 6000 
km s$^{-1}$.

This paper is organized as follows: in Section 2, we describe the
observational sample; in Section 3, we outline the procedure
used to analyze the galaxy photometry, and assess the quality of our
data. The photometric catalogue is presented in Section 4, while a
summary of our main results concludes in Section 5.

\section{The Data}
\label{obs}

\subsection{The Sample}
\label{sample}

A detailed description of the ENEAR survey can be found 
in da Costa et al. (2000a) who describe the sample selection,
properties and completeness, so we only present a
brief overview here.  The ENEAR sample was drawn from an all--sky source
catalogue of galaxies of all types by selecting objects
brighter than $m_{B(0)} $= 14.5 mag and  with morphological types T$\leq$
-2 (following the morphological classifications of Lauberts and
Valentijn 1989) and with radial velocities $V_r \leq 7000 $ \kms; this sample
will be referred to as ENEARm. 
The distances were estimated with a template \dnsig relation (Bernardi et
al. 2002a, 2002b) using the subsample of galaxies in 28 groups and 
clusters for which we use the name ENEARc. The galaxies were
assigned to 23 of them by applying an objective group-finding algorithm to the
source catalogue (which contains galaxies of all morphological types).  We also
added five additional well-studied clusters.  
This ENEARc subsample also contains 134 objects that are either fainter than 
$m_{B(0)} $= 14.5 or with $V_r > 7000 $ \kms.  This means that while there is a considerable
overlap between ENEARm and ENEARc neither sample contains the other in its 
entirety as a subset.  In addition, while for the ENEARc sample 
(Bernardi et al. 2002a, 2002b), we 
combined our measurements with those of the literature, this paper reports
only our new measurements.

The photometric data presented here consist of
1332 galaxies: 1104 objects belong to ENEARm, of which 201 are galaxies 
in clusters contained in the ENEARc sample. There are an additional 134 
galaxies (with $m_{B(0)} > 14.5 $ or with $V_r > 7000 $ \kms) 
belonging to ENEARc, as explained above and  
finally 94 galaxies that are contained in neither
sample.  Most of the latter are serendipitous early-type galaxies which lie in the
same CCD frame as an observed programme galaxy, and generally have  $V_r > 7000 $ \kms.  

\subsection{Observations}

The R$_C$-band (Kron-Cousins) photometry reported in this paper was
obtained over 100 photometric or partially photometric nights out of a
total of 177, using several telescopes over
various observing runs in the period 1987 -- 1999. The following
telescopes were employed: the  Danish (hereafter DK) 1.54m and Dutch 0.9m 
telescopes at the European
Southern Observatory (ESO); the 0.9m telescope at Cerro Tololo
Interamerican Observatory (CTIO); the 0.61m and 1.3m telescopes at
Fred Lawrence Whipple Observatory (FLWO) and the 1.3m telescope at
MDM (formerly the Michigan--Dartmouth--MIT) Observatory.

The basic information for each
run is summarized in Table~\ref{tab:runs} where we list: in column (1)
the identification code of the run; in column (2) the observing date; in
column (3) the number of the total/photometric nights and in column (4) 
the corresponding reference number of the instrumental setup, which is
described in Table~\ref{tab:setups}. 

A total of 12 different setups were used, corresponding to
different telescope/detector combinations and are described in
Table~\ref{tab:setups} which gives: in column (1) the setup reference
number; in column (2) the observatory and telescope identification;
in columns (3) and (4) the total number of images observed in the R$_C$ band
($N_m$) in 
that setup and the number of repeated images ($N_r$),
which are used later as calibrators to homogenize our observations; and in
columns (5) to (9) some characteristics of the detectors:
identification, size, pixel scale, gain and read--out noise.  
It is important to mention that setups 3, 4 and 5 correspond to the 
DK 1.5m telescope with DFOSC (Danish Faint Object Spectrograph and Camera) 
and the CCDs identifications are reported in column (5).  

Exposure times
 varied from 120 to 600 seconds depending on the telescope
and the brightness of the galaxy. A total of 2339  images  obtained under 
photometric conditions
were analyzed. Of these, 2121 were taken with the R$_C$ filter
 and 218 with B.  The latter sample, comprising 178 galaxies, 
will be discussed in a separate paper (Alonso et al. 2003a).
The final sample consisting of 1332 galaxies has been constructed
after discarding 129 galaxies for a variety of reasons (e.g. 
superimposed objects; crowded fields) and about 50 galaxies
observed too close to the edge of the CCD or with low signal-to-noise.

Finally, it is important to point out that we have a total of 257 objects with
multiple observations using either the same or different setups. 
Given the long duration of the program and the large number of setups 
involved, these repeated observations are of paramount
importance to ensure the overall uniformity of our data
and were used to make our measurements of
photometric parameters internally consistent and to estimate their
errors.

\subsection{Data reduction and calibration}
\label{landolt}

All images were trimmed, bias subtracted and flat-fielded using standard
IRAF\footnote{IRAF (Image Reduction and Analysis Facility) is
distributed by the National Optical Astronomy Observatories.} routines.
The bias, dome and sky flats, obtained over one or more nights of a
given run, were median combined and
both sky flats and dome flats were used to investigate the quality of the
flat-fielding.
The uncertainty in the residual large--scale response of the CCD
measured using the sky-flats was found to be less than 1\%.  No
corrections for dark current were required as it was determined to be
negligible for all  detectors.

\begin{figure}
\mbox{\psfig{figure=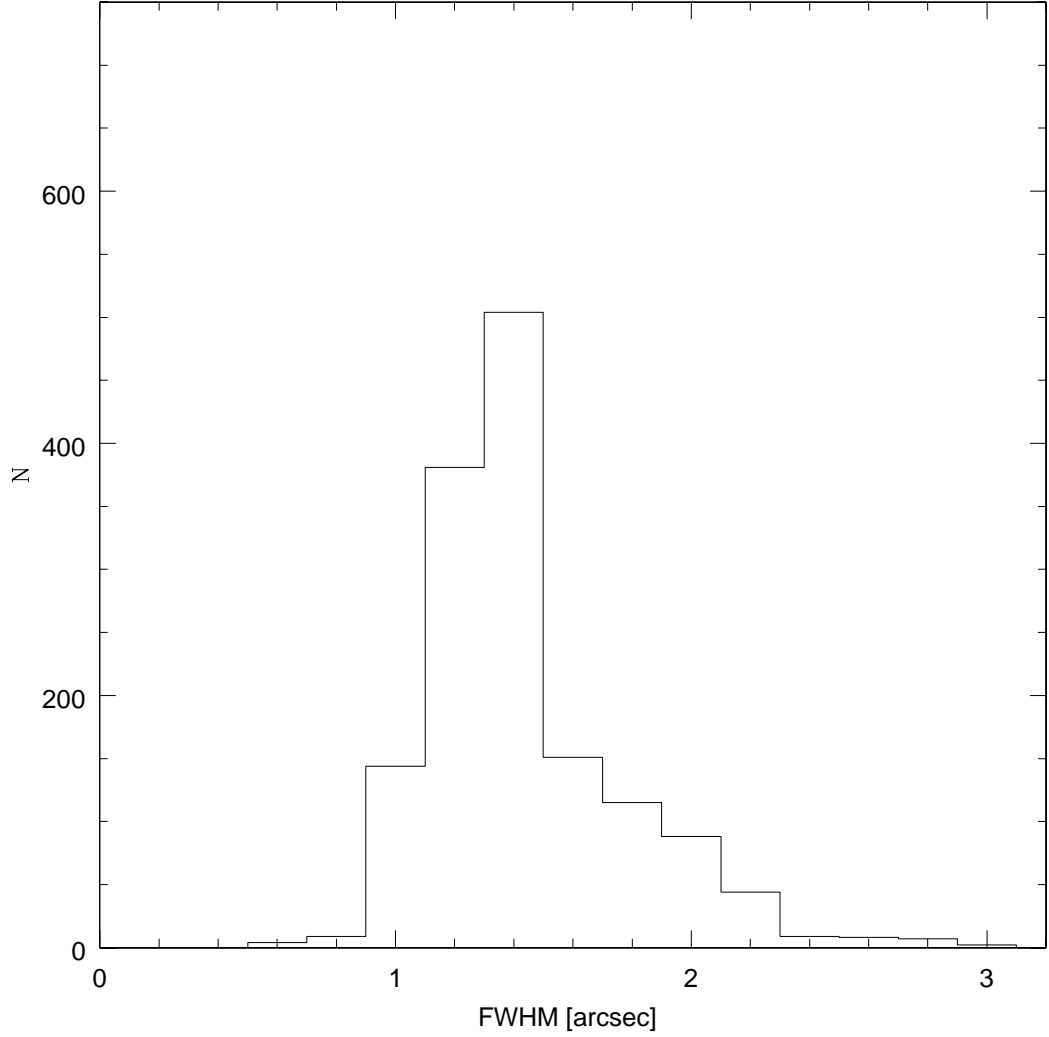,height=15truecm }}
\caption{  Distribution of the PSF FWHM values in arc seconds, as measured
from stars observed in the same images as the programmed galaxies. 
The median seeing is $\sim$ 1.39~arcsec. The distribution is skewed to
high values, which are mostly caused by observations made under 
unfavorable conditions.}
\label{fig:fwhm}
\end{figure}

The photometric calibration relied on observations of
Landolt (1983, 1992) standard stars in the Kron-Cousins 
R$_C$ band and, depending on
the run, in a second passband, generally V, but sometimes in B and I$_C$, to
obtain color corrected solutions. In
general, standards were observed at $\sim 1-2$ hour intervals,
covering a wide range of colors and airmass throughout the night.  
Typically, about 30 stars were observed during each night.  We followed
the reduction procedure of Haynes et al. (1999b): instrumental
magnitudes for the stars were obtained using a circular
aperture large enough to measure the total flux without significantly
increasing the error due to sky--noise. A suitable aperture was chosen
for each run depending on the observed stellar fields, where we tried 
to minimize the
rms obtained for the photometric solution in that run. Typical values
of the aperture radii were $\sim 6$ arcsec. For the median seeing of 1.39 
arcsec (see Figure~\ref{fig:fwhm}) this is about 9 times larger than the
point-spread function (hereafter PSF FWHM).  The sky level was
determined as the median value of counts measured within an annulus
of about 16 arcsec radius centered on the standard star, being far enough that
the contribution from light in the stellar wings should be negligible.

The photometric solution was obtained for each night of a given run 
and mean values of the zero point and color term were taken for that run.
Once they were fixed, the extinction coefficients were determined nightly 
for each run. Nights were considered photometric if the dispersion 
between the standard magnitudes and those obtained from the
fits was $\lsim 0.05 $~mag. Using this criterion
94 out of the 177 nights allocated to the project were
considered photometric.  Six partially photometric
nights were
added after discarding the portions observed under unfavorable
conditions. Since no colors are available for most of
our galaxies, we have assumed mean colors of
early-type galaxies as being $<$B--R$_C>=1.48$, $<$V--R$_C>=0.56$, and
$<$R$_C$--I$_C>$=0.70~mag (Frei \& Gunn 1994; Fukugita, Shimasaku \& Ichikawa
1995). From the color terms for our nightly photometric solutions, which are
typically 0.01, the uncertainty introduced by this assumption is of about 
0.05~mag.

The distribution of the observed PSF FWHM, as measured
from stars observed under photometric conditions off the same images as
the program galaxies is shown in
Figure~\ref{fig:fwhm}. The median value of the distribution is  $\sim$
1.39 arcsec but shows a tail extending to large values. 
 Therefore, since the seeing FWHM is not always negligible compared 
to the sizes of the galaxies in the sample (see Section~\ref{fitproc}), all
measurements of the photometric parameters were made on light profiles 
corrected for seeing.

\section{Surface Brightness Profiles}

  The measurement of photometric and structural parameters of the galaxies
employed
the GALPHOT package originally developed for spiral galaxies
(Haynes \etal 1999b and references therein). Background estimates
were obtained from ``sky boxes'' placed in regions around
the galaxy that are free of bright stars, but far enough not to be 
contaminated
by light coming from the outer parts of the galaxy or other resolved
objects in the frame. For each ``sky box'' the
mean intensity was computed after automatically masking out faint stars
and galaxies within the box. The final sky value was computed as the
mean of all values measured in the ``sky boxes'', since the images were 
sufficiently flat. This average sky value
was then  subtracted from the image.  The typical scatter
of the mean sky intensity  measured in each ``sky box'' was $\lsim
0.5$\%.

Finally, prior to conducting the  surface photometry, a
rectangular region about twice the size of the galaxy image was marked
and cosmic rays and stars outside this box were automatically masked.
Standard IRAF routines were utilized to identify the different objects
in the images above a given threshold and the classification of them
was based on its roundness and sharpness.
Stars which are within the rectangular region 
were not marked automatically to avoid eliminating important parts of the 
galaxy. Any remaining undesirable features both inside and outside the box
were masked interactively.  Masked stars and cosmic rays were not considered
in calculations of the flux within different apertures.

We measured the surface brightness profiles of all galaxies using
both circular and elliptical apertures. In both cases, the photometric
center of the galaxy relied on the ellipse fitting method of
Jedrzejewski (1987), from which one quantifies the shape and orientation
of the  galaxies, and the deviations of the isophotes from perfect ellipses. 
In this paper we only
consider the profiles derived from circular  apertures, and
defer the discussion of the photometry using elliptical profiles to
another paper (Alonso et al. 2003b).

\subsection{Circular Averaged Aperture Profiles}

The circularly averaged light profiles were measured using 1 pixel
steps starting from an innermost radius  of 1 pixel (ranging from
0.38 to 0.65 arcsec depending on the scale) 
to an outer radius  $r_{max}$, where the light profile
counts drop below  the
1-$\sigma$ level of the sky background.  The center for the
aperture photometry was assumed to be identical to the smallest
ellipse derived from the two-dimensional isophotal fit (Alonso \etal 2003b).
The instrumental values obtained for the surface brightness
and magnitudes were calibrated using the photometric solution 
described in Section~\ref{landolt}.  The surface brightness 
profiles were also corrected by galactic extinction, the
$K$--correction (Davis et al. 1985) and the $ (1+z)^4$ cosmological effect.  
The galactic extinction
in the R$_C$-band was estimated as A$_R$ = 0.58 A$_B$ (Seaton 1979) where A$_B$
was based on the maps of Burstein \& Heiles (1984).  The comparison of these
extinction estimates with those of Schlegel, Finkbeiner \& Davis (1998) gives
a mean difference of -0.07$\pm$0.05, or a mean difference in A$_R$ of about
0.04.  For the observed galaxies, the galactic extinction correction is 
A$_R$ $\lsim 0.13$ mag.  

The primary goal of our imaging survey has been to determine photometric
quantities that can be used in the empirical distance
relations, such as the \dnsig (Dressler et al 1987), 
to map the peculiar velocity field.  The characteristic angular diameter 
$D_n$, as
originally defined by the 7S, is the circular diameter within which the
corrected average surface brightness of the galaxy 
is equal to 20.75 \bri in the B band.  
The choice of isophotal level where \dn is defined is influenced 
by the presence of the disk.  Using a faint isophotal level, especially
in lenticular galaxies, may include a contribution from the disk component
which can have different dynamical properties.  At
brighter isophotal levels the values are sensitive to seeing,
especially for the more distant galaxies.  If we follow Dressler (1987) 
who measures $D_n$ at 19.75 \bri in the B band for lenticular galaxies,
we find very small values which are strongly affected by seeing.  Moreover, in
some studies (\eg J\o rgensen, Franx \&   Kj\ae rgaard 1995, hereafter JFK
 and Lucey \etal 1991) the
diameters have been measured at the same level independently of
morphological type.  Consequently in order to measure \dn, we
adopt the isophotal level of $\mu_{R_C} = 19.25$ \bri in the
R$_C$-band regardless of the morphological type.  This
corresponds to
the 7S B band definition assuming a mean color of
 $<\mu_B-\mu_{R_C}>$ = 1.5~\bri, which we obtained from
galaxies in our sample observed in both bands.

The effective radius, \re is the radius of the isophote
that encloses half the luminosity of the galaxy bulge and \mue is
the mean surface brightness within \re. As we explain below, all these
parameters are corrected for seeing effects (Saglia \etal 1993).
The circularly averaged growth curves for most elliptical galaxies do not 
differ significantly from those derived from elliptical isophotes (Saglia
 \etal 1993). Both growth curves are equivalent when the surface brightness
profile of a galaxy is described by a pure $r^{1/4}$ law and the ellipticity is
constant. Even though most of the surface brightness profiles of 
early-type galaxies are a
combination of bulge and disk components, the difference introduced
by using circular apertures is small.  In the cases of galaxies showing
flattened bulges or an evident disk component, especially if they are
seen edge-on, the values of the structural parameters are unreliable.
These problems arise for objects with ellipticities $>0.6$
(see Alonso et al. 2003b)
which represent only a small fraction of the total (\lsim  4\%).

\subsubsection{Quality of the profiles}
\label {surf}

The internal accuracy was estimated using galaxies
with more than one observation. Our sample includes 257 objects which have
multiple observations obtained using either the same or different setups.
The number of repeated observations ranges from 2 to 7: 206 objects were 
observed twice; 33 objects, 3 times; 10 objects, 4 times; 5 objects, 5 times;
only 1 object, 6 times and 2 objects, 7 times. These repeat observations 
can be of three different types:
galaxies observed more than once during the same
night; galaxies observed on different nights but using the same
observational setup; galaxies observed with
different setups. By splitting the comparisons in this way it is
possible to evaluate
the stability of the photometric solution over a night, quantify the
impact of seeing variations and estimate our internal errors.  Comparisons of
galaxies observed on different nights are also useful
since they reflect the more general cases of combining observations
under different atmospheric conditions without introducing issues
related to color-terms, field-of-view and other instrument and telescope
dependent quantities.  Finally, the comparisons between galaxies observed
with  different setups allow an evaluation of the zero-point
calibration providing an additional check
on the accuracy of our photometric solutions.

When making the comparison, we used the convention of performing the
differences between  "older minus newer" measurements.  For instance,
we compared a measurement taken at ESO-601 with all the other ESO runs.  
Then we compared ESO-603 measurements  with later runs (e.g. ESO-604, 
ESO-605, ESO-606, etc. but not with ESO-601). The 
profiles were compared point by point and the mean
weighted differences were computed using a range of radii ($r_{min}$
and $r_{max}$, see above).  This radial interval was chosen to avoid
regions where the smearing of light due to seeing is significant,
while the outer radius was chosen to minimize the contamination from
other objects in the field.  
In general galaxies that present differences at large radii are either 
located in
crowded stellar fields or have relatively bright stars superposed,
so that an accurate mean sky level is difficult to measure.
Other possible causes of large differences are
residual contamination from nearby galaxies;
and extended galaxies reaching the edge of the
detector. At inner radii the center positions of the isophotes are very 
uncertain 
when there is light contamination by background stars, absorption features or
when there is more than one surface brightness peak (probably dumb-bell 
systems).  
In the final comparisons we excluded objects that were flagged as
contaminated or with uncertain sky subtraction.
 The complete set of plots with the
differences in the surface
brightness profiles  and a detailed description of the comparisons were
 presented by Bernardi (1999).

These results are summarized in Table~\ref{tab:compstatint} which
gives: in column (1) the comparison set considered; in column (2)
the number of different profiles being compared;
 in columns (3) the mean difference in surface
brightness and error; and in column (4) the scatter.
These values are consistent with those obtained from a similar
comparison of profiles determined by the ellipse fitting procedure.
The observed scatter in the comparisons are consistent with the
accuracy of the photometric solutions ($\delta m \sim 0.05~\rm{mag}$),
giving a zero point estimate of about 0.037.  
The color term contribution to the surface brightness is, in general, 
smaller than 0.06 \bri.  So, the larger scatter found in the comparison among 
different setups may reflect the uncertainties in the photometric
calibration introduced by the zero point and 
different instrumental color terms.

\subsection {Fitting procedure}
\label {fitproc}

\subsubsection{Seeing corrections and Photometric Parameters}

As our sample includes both elliptical and lenticular galaxies,
we examine how the parameters involved in the scaling
relations should be determined in the case of a two-component
system.  Following Saglia \etal (1997),  the
surface brightness profiles within circular apertures
were fit using a two-component model
comprising a bulge ($r^{1/4}$) and an exponential disk, convolved with a
PSF FWHM (Saglia et al. 1993) to correct for seeing effects. 
The programs to make this seeing corrected profile decomposition
were kindly provided by R. Saglia. The Fourier
transform of the PSF FWHM is assumed to be $\approx
\exp[{-(kb)^\gamma}]$, where $b$ is a scaling parameter
and $\gamma$  is a parameter that describes the shape of the PSF FWHM.  In
general,  $\gamma$ varies between 1.3 for profiles with extended
faint parts; and 2 for profiles with a sharp cut.  In our convolutions
we assume $\gamma $= 1.6.  The seeing can be significant out
to a radius few times larger than the PSF FWHM, and thus lead to 
substantial errors when the ratio $r_e$/(PSF FWHM)/2 is $\sim 1$. 
Since the median seeing of our observations was $\sim 1.4$ arcsec and 
the median effective radius of the galaxies in the ENEARm sample is $\sim 22$
arcsec (as we will see below), seeing corrections are significant
for $\lsim 25$\% of the galaxies. However, for faint ENEARc galaxies in 
clusters the seeing
correction is important for at least half the observed galaxies.

Saglia et al. (1993) show that  the presence of a disk component
shifts galaxies away from the Fundamental Plane for elliptical
galaxies. These deviations also correlate with the galaxy's
ellipticity, and will be discussed by Alonso et al. (2003b). Thus, in
the present sample we use a two-component fit to
the surface brightness profile, in contrast to some previous
works (e.g., Burstein \etal 1987; JFK; Scodeggio
\etal 1998) which used only a one-component fit to determine the
photometric parameters used in the distance indicator.
Saglia \etal (1997) also found
that ignoring the disk component could bias the results leading to
errors in $r_e$ of $\sim 20$\%. However,
by combining some of the galaxy parameters, the disk's contribution
can be sufficiently small so that its effect is canceled.
Smith \etal (1997) employed this procedure by combining
the effective
radius with the mean surface brightness (using \lre -- 0.35 \mue) which is
little affected by the presence of a disk, since errors in
$r_e$ and \mue are correlated. Our photometric data show
no dependence of this quantity on the D/B ratio.  Thus, using only the 
bulge component of lenticulars should not produce a tighter scaling 
relation for these galaxies, which is supported by the lack of any 
correlation in the \dnsig relation as a
function of the $D/B$ ratio (Bernardi \etal 2002a).

Each galaxy profile in our sample is fit
three times using: (i) a pure r$^{1/4}$ law ($D/B=0$); (ii) an
exponential disk profile ($B/D=0$); (iii) the sum of a bulge and a
disk component. In all cases these profiles are convolved with the PSF FWHM
as described above.  The $\chi^2$ is measured in the standard way, using
the bulge effective radius ($r_e$); the disk exponential scale
length ($\alpha$); the disk-to-bulge ratio ($D/B$), and the seeing FWHM 
as fit parameters. The weights are the statistical errors in the values 
of the surface brightness at each radius.  For each fit one can either 
assume a fixed mean sky value using ``sky boxes'' or allow it to be a fit 
parameter. Therefore, the method produces six sets of parameters.
The $\chi^2$ is computed for each of the six
fits, and the one with the smallest reduced $\chi^2$ is chosen as the
best fit. Visual examination shows
that in most cases the fit with the smallest $\chi^2$
is reliable. However, in less than 1\% of the cases we
find that the best fit measured using $\chi^2$ lead to artificially
large disks. In these cases we use the results obtained from
fits with slightly larger $\chi^2$ but which are more consistent 
with the data profiles.

\begin{figure}
\vspace{5cm}
\centering
\mbox{\psfig{figure=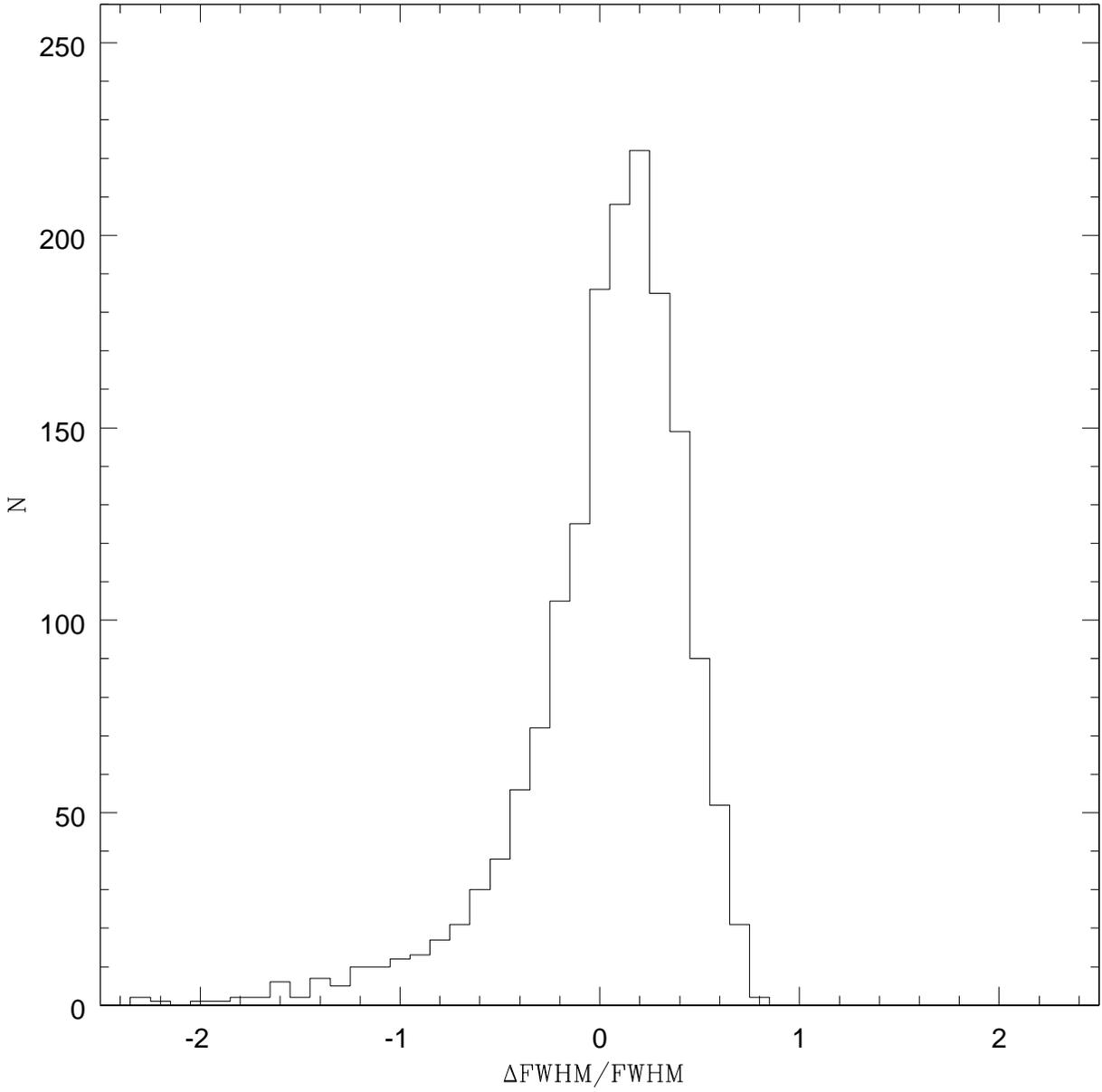,height=13truecm,bbllx=2truecm,bblly=5truecm,bburx=19truecm,bbury=20truecm}}
\caption{The distribution of the $\Delta$ FWHM / FWHM (measured), where \break
$\Delta$ FWHM =  $\rm{FWHM (measured) - FWHM (fit)}$.}
\label{fig:seeingdiff}
\end{figure}

The FWHM as determined from
the fits agree reasonably well with that measured from stellar
profiles on the same galaxy frames. In Figure~\ref{fig:seeingdiff}
we plot the comparison between the PSF FWHM measured both ways.
The distribution peaks around zero but is skewed towards
negative values indicating that, in general, the fitting procedure
overestimates the seeing.
The cases presenting large differences are due to poor fits, generally of 
galaxies observed in early runs using smaller CCDs or observations taken
under poor conditions.

\begin{figure}
\vspace{5cm}
\centering
\mbox{\psfig{figure=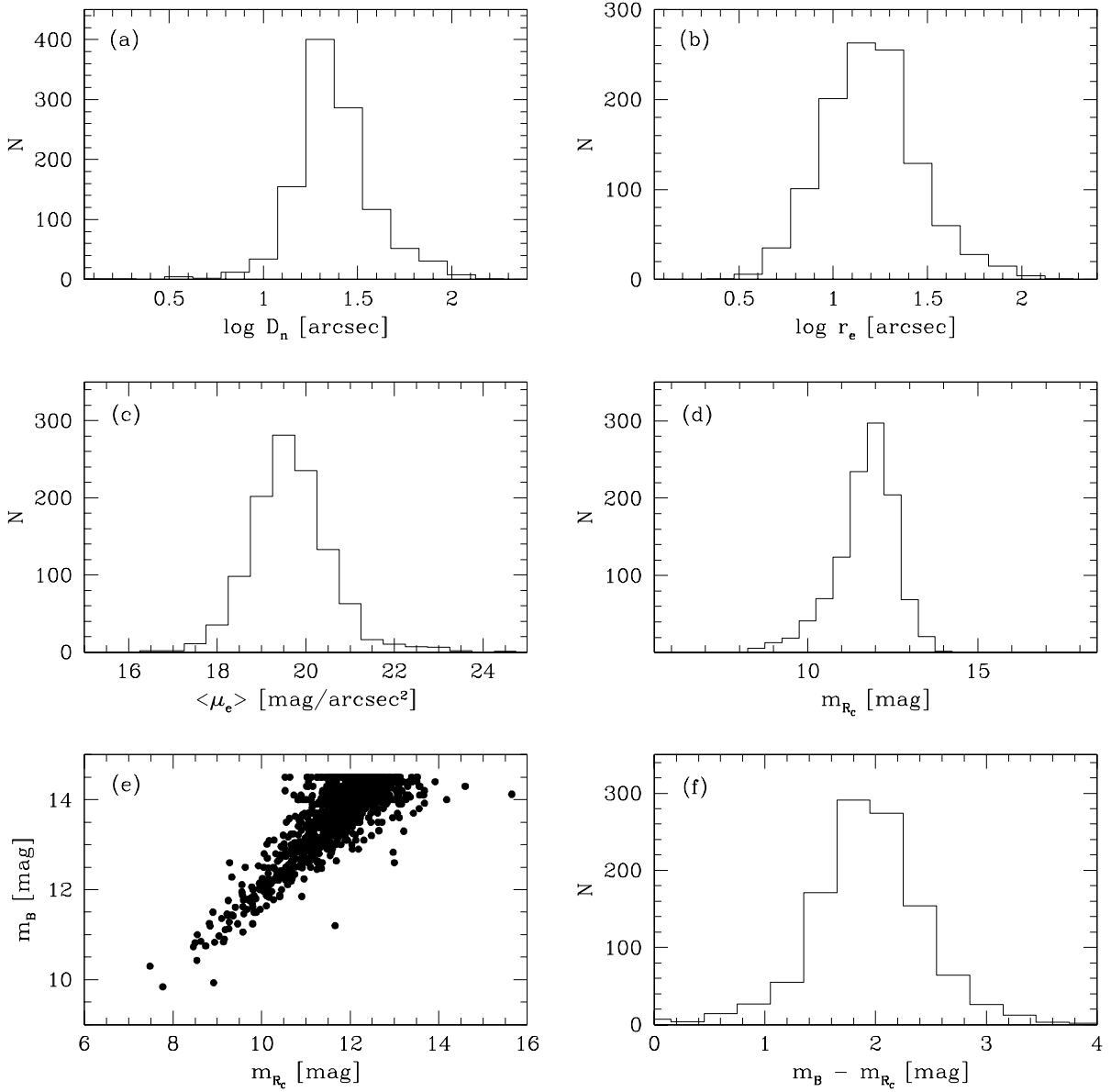,height=13truecm,bbllx=2truecm,bblly=5truecm,bburx=19truecm,bbury=20truecm }}
\caption{Distribution of the photometric parameters derived for
the ENEARm sample. The $D_n$ parameter is shown in arc seconds so that it
may be directly compared to $r_e$. The panel (e) shows the
B-band versus the R$_C$-band total magnitude and the panel (f) the
distribution of the color B-R$_C$.}
\label{fig:histphotenearf}
\end{figure}

The seeing corrections make the overall galaxy light profile
brighter and steeper in the innermost regions
yielding structural parameters with larger $D_n$, brighter \mue 
and smaller $r_e$ when compared to the uncorrected measures.
The correction for $D_n$ is only important for galaxies with
$D_n \sim 10$ arcsec, which
in the case of ENEARm affects only very few galaxies (see
Figure~\ref{fig:histphotenearf}). For values of
$D_n>$ 10 arcsec, $D_n$ only increases by at most about 2\%.  However, the seeing 
corrections do become important for the ENEARc sample of galaxies in clusters, where the
corrections can be as large as 20\%.

\subsubsection{The fit quality parameter Q}
\label{results}

We assign a quality parameter $Q$ to the profile fits adopting the method of
Saglia \etal (1997) who use the following criteria: (i) the extent of the
profile ($r_{max}$) relative to $r_e$; (ii) the influence of seeing (the
PSF FWHM compared to $r_e$); (iii) the value of the integrated galaxy S/N;
(iv) the galaxy's surface brightness relative to the sky; (v) the uncertainty in
the sky determination; (vi) the fraction of the total light derived by
extrapolating the profile beyond $r_{max}$ used in the total magnitude
$m_{R_C}$; (vii) the reduced $\chi^2$ goodness-of-fit. The quality parameter
ranges from excellent ($Q$ = 1) to poor ($Q$ = 3).

Monte-Carlo modeling
of the EFAR sample enabled Saglia \etal (1997) to relate
$Q$ to the errors in the
photometric parameters. Fits with $Q$ = 1 led to the following errors:
1) $m_{R_C} \lsim 0.05$ mag; 2) \lre $\lsim$ 0.04; 3) \ldn and FP defined
as \lre -- 0.30\mue $\lsim$ 0.005. For $Q=2$ the respective errors are
$\lsim 0.15$ mag; $\sim$ 0.11 and  $\lsim$ 0.01. For $Q=3$, the errors are
larger than these values.  Below we calibrate
the errors in the ENEAR data and we find that our errors are comparable
with those of Saglia \etal (1997).

\begin{figure}
\vspace{5cm}
\centering
\mbox{\psfig{figure=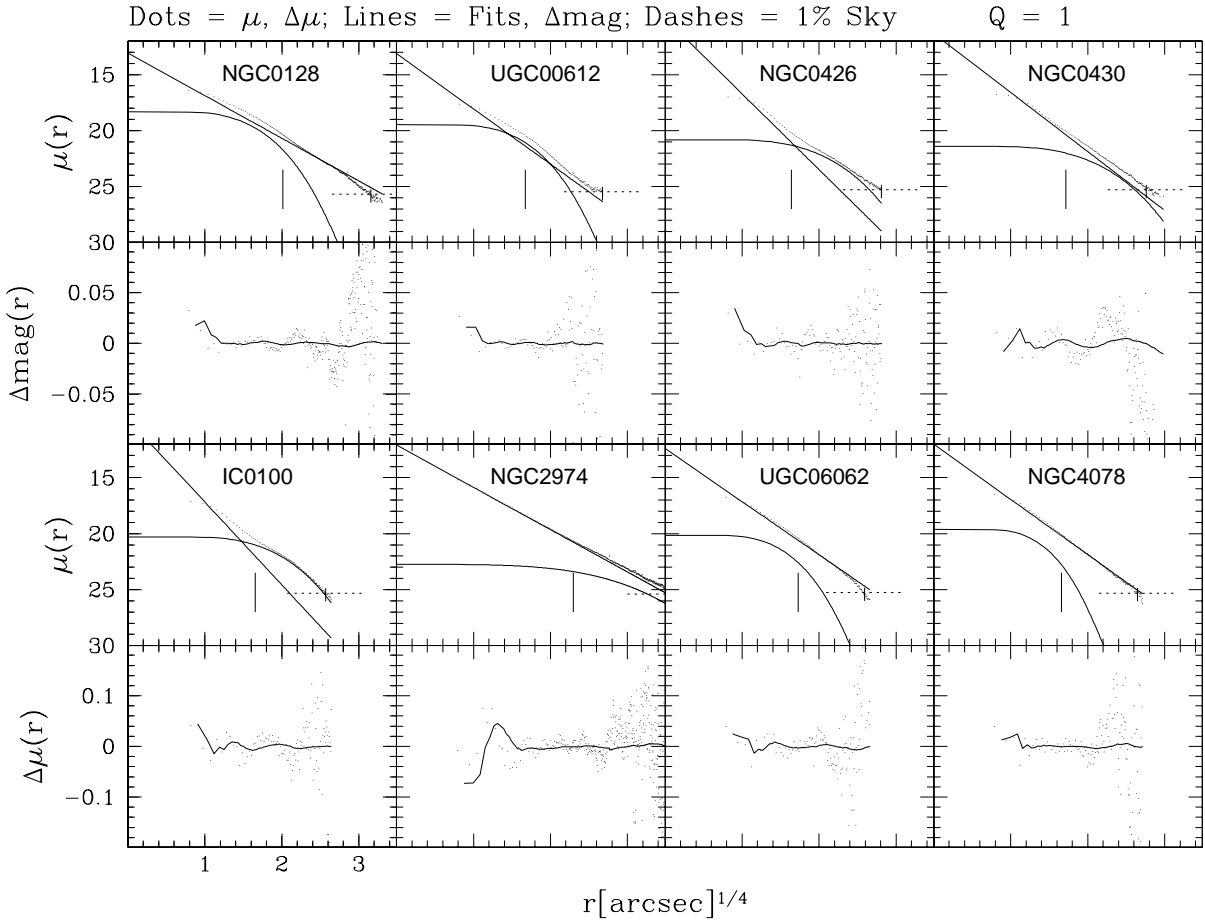,height=12truecm,bbllx=1truecm,bblly=5truecm,bburx=20truecm,bbury=22truecm,angle=-90 }}
\caption{Examples of the light profile fitting of excellent quality (Q = 1).
For each galaxy there are two plots showing: in the upper panel, the observed
light profile (small dots) and the best-fit (solid lines) as a function of
$r^{1/4}$. The larger vertical line marks the derived value of $r_e$, while
the smaller shows the maximum extent of the profile, $r_{max}$. The horizontal
dashed line is the intensity corresponding to 1\% of the sky.  The lower panel
shows the differences $\Delta \mu $ = $\mu_{obs}(r)$ - $\mu_{fit}(r)$ 
(small dots) and $\Delta m $ = $m_{obs}(< r)$ - $m_{fit}( < r)$ (solid line).}
\label{fig:fitQ1}
\end{figure}

\begin{figure}
\vspace{5cm}
\centering
\mbox{\psfig{figure=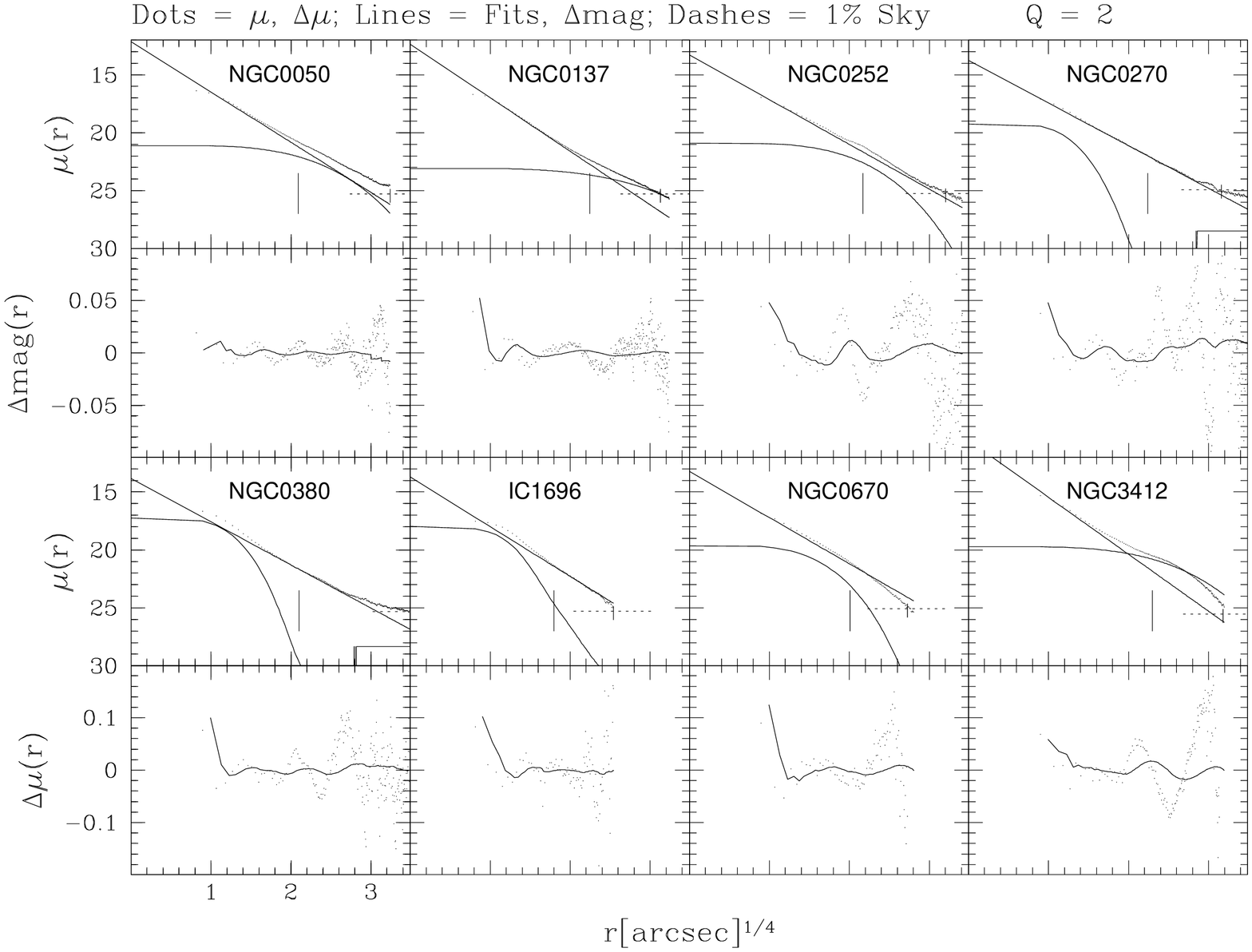,height=12truecm,bbllx=1truecm,bblly=5truecm,bburx=20truecm,bbury=22truecm,angle=-90 }}
\caption{Examples of the light profile fitting of moderate quality (Q = 2).
The two panels for each galaxy show the observed light profile, the best fit
and differences in surface brightness and magnitudes as in 
Figure~\ref{fig:fitQ1}.}
\label{fig:fitQ2}
\end{figure}

\begin{figure}
\vspace{5cm}
\centering
\mbox{\psfig{figure=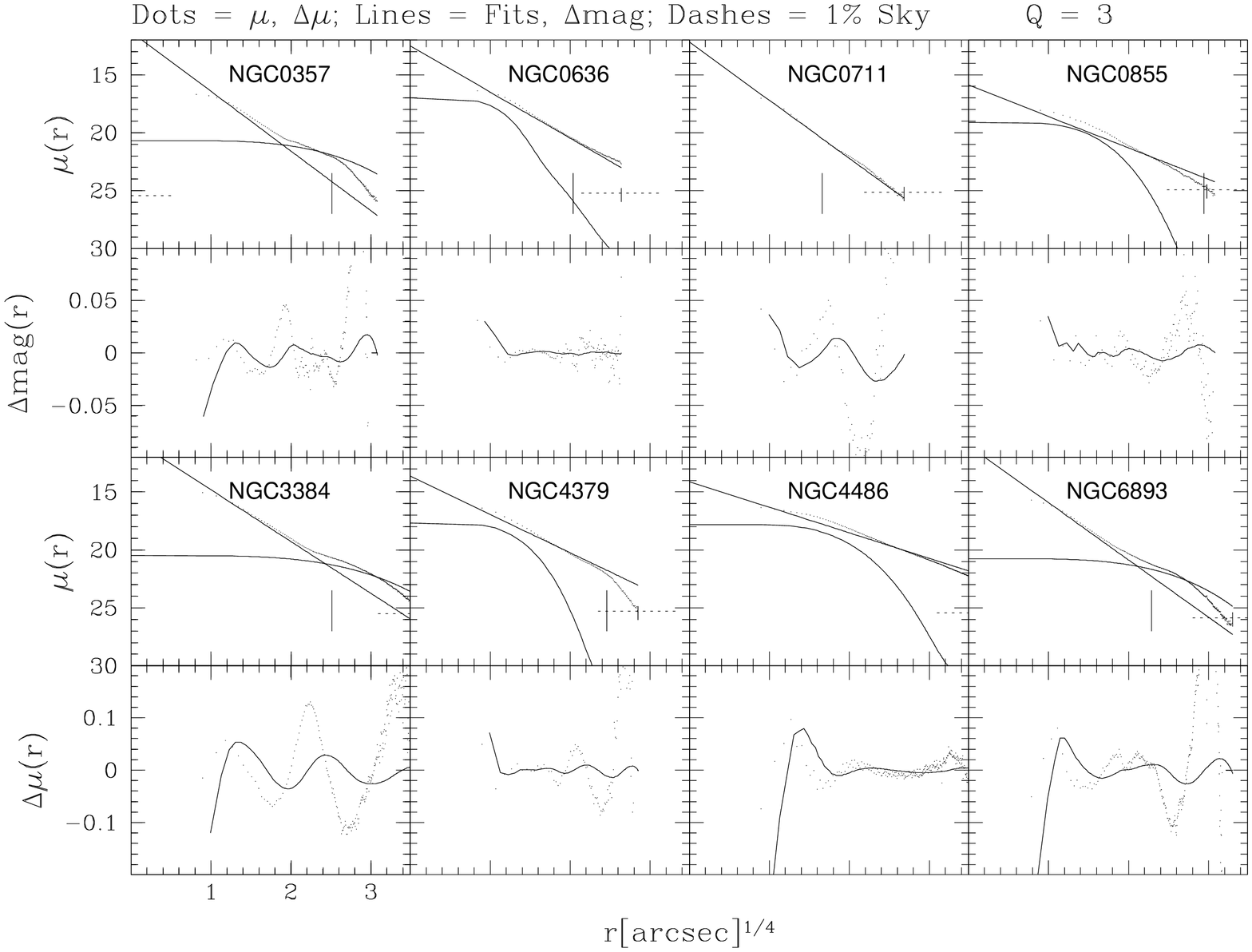,height=12truecm,bbllx=1truecm,bblly=5truecm,bburx=20truecm,bbury=22truecm,angle=-90 }}
\caption{Examples of the light profile fitting of poor quality (Q = 3).
The two panels for each galaxy show the observed light profile, the best fit
and differences in surface brightness and magnitudes as in 
Figure~\ref{fig:fitQ1}.}
\label{fig:fitQ3}
\end{figure}

Fit results of different quality are compared with the observed
light profiles in Figures~\ref{fig:fitQ1},
\ref{fig:fitQ2} and~\ref{fig:fitQ3}.
For each galaxy these figures show two panels: in the upper
one, the observed light profile (small dots) and the fitted bulge and 
disc profiles (two solid lines)
as a function of $r^{1/4}$ and in the lower panel, the difference in
surface brightness $\Delta \mu = \mu_{obs}(r) - \mu_{fit}(r)$ (small dots,
scale -0.2 to 0.2 \bri) and the integrated magnitudes \break
$\Delta m = m_{obs}(< r)$ - $m_{fit}(< r)$ 
(solid line, scale -0.1 to 0.1 mag) 
derived from the observed and the fit profiles. The two differences are
plot together but, for instance, the lower panel of NGC0128 
(in Figure ~\ref{fig:fitQ1})
is showing the scale and label corresponding to $\Delta m $ and the lower
panel of IC0100, the scale and label of $\Delta \mu$.  
The usefulness of $Q$ to
qualify the profile fits is evident in those cases of poor fits ($Q$ = 3),
 where the inspection of the galaxy images shows the presence of spiral
arms and/or bars, generally indicative of misclassifications
in the original catalogues. Cases showing large deviations from the fits are
indicated  in Table~\ref{tab:data}.

We have adopted the
definitions of Saglia
\etal (1997) for all $Q$ parameters except for $\chi^2$ which
required renormalization to the S/N of our data in order to employ the results
of their simulations. While $Q$ indicates the quality of the fit and the
resulting photometric parameters, it was only used to assess the profile fits.
The final errors were estimated using galaxies  with multiple observations
and the accuracy of these errors were assessed using
comparisons with other authors (see Section~\ref{errors}).

\subsubsection{Results of $Q$ measures for the ENEAR galaxies}
\label{results}

The analysis of the profile fits for galaxies in ENEARm shows that 
$\sim$ 12\% are
well described by the pure $r^{1/4}$ law while 87\% are best fit by the
two-component model, 78\% of these cases having $D/B < 1$. Less than 
1\% of the galaxies
have $D/B > 20$, and these usually show signs of nearby companions, spiral
arms, etc. (see Table~\ref{tab:data}). For $\sim$ 26\% of the 
observed galaxies
our results suggest that the morphological classifications in the source 
catalogues (all of which used photographic material) must be revised, as
in most of these cases the CCDs with their larger dynamic range 
show the presence
of bars and or spiral arms. Both the ENEARm and ENEARc
samples include some disk-dominated galaxies. 

For the ENEARm sample, 60\% of the fits have $Q = 1$, 22\% $Q = 2$, and
18\% $Q = 3$. Most $Q = 3$ fits have relatively bright outer isophotal levels and
truncated profiles. This not only impacts the $Q$s of the sky correction
but also requires large radial extrapolations when computing the total
magnitudes. The remaining poor fits ($\lsim$ 30\%) are caused by additional
features in the light profiles as already described. For the ENEARc sample we
find a higher percentage, 18\%, of pure $r^{1/4}$ profiles. About 47\% of
the galaxies have $Q = 1$ fits, 27\% $Q = 2$, and 26\% $Q = 3$.

\begin{figure}
\vspace{4cm}
\centering
\mbox{\psfig{figure=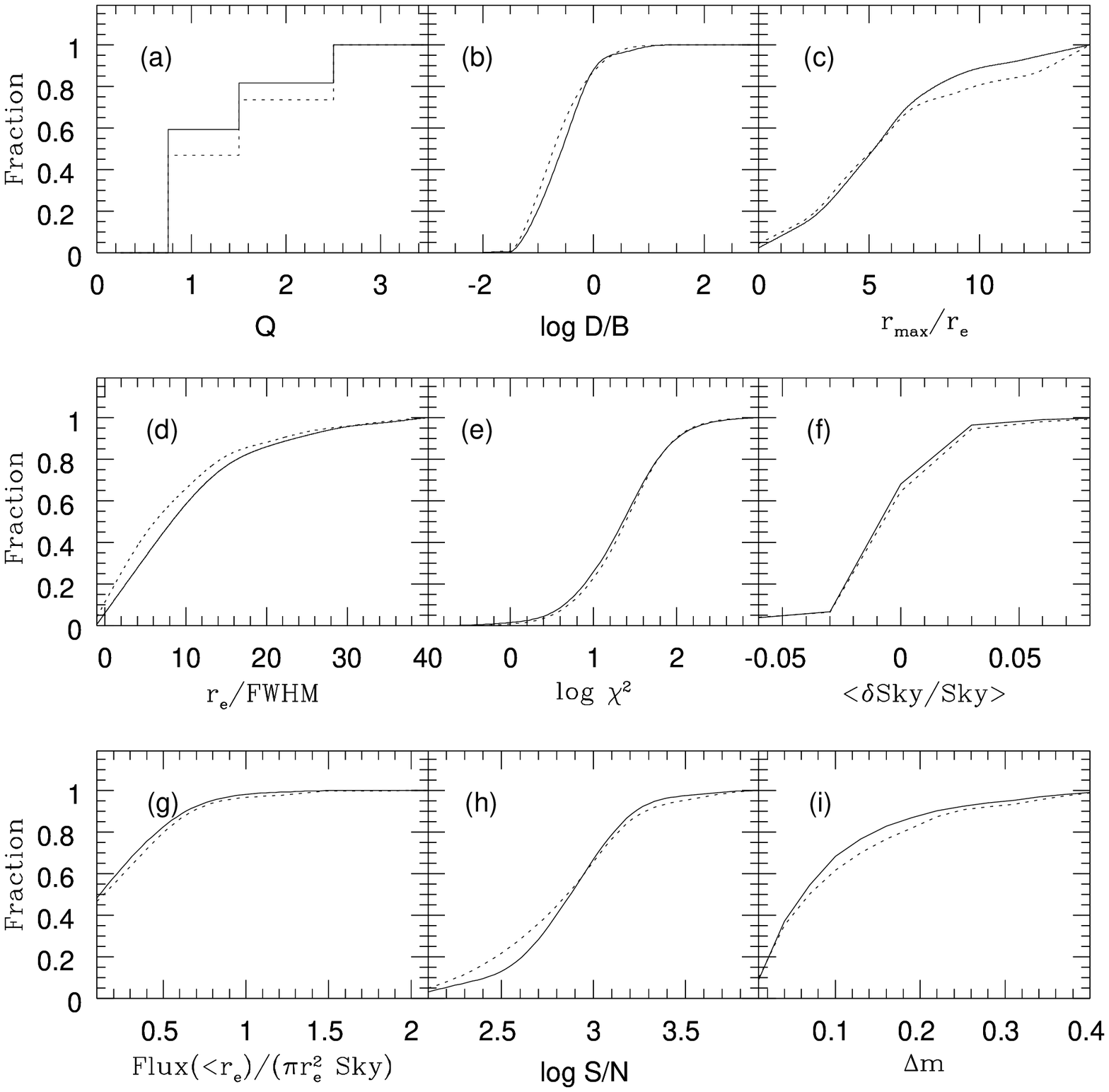,height=13truecm,bbllx=2truecm,bblly=5truecm,bburx=19truecm,bbury=20truecm }}
\caption{{\small  Cumulative distribution of the parameters which are obtained
from the surface brightness profile fits. The full line represents the
ENEARm sample and the dashed line ENEARc. The panels show the following
parameters: in panel (a) the  fit quality $Q$;
(b) the logarithm of the disk-to-bulge ratio ($\log {D/B}$ = -1 for simple 
r$^{1/4}$ fits); (c) the maximum extent of the profile ($r_{max}$) compared to
$r_e$; (d) the ratio between $r_e$ and the fitted PSF FWHM; (e) the
logarithm of the reduced $\chi^2$
of the fit; (f) the average sky correction, whenever applicable;
(g) the ratio between the flux of the galaxy and the sky within $r_e$; (h) the
logarithm of the total signal-to-noise ratio of the profiles; and (i)
the amount of extrapolation used to compute the total magnitudes.}}
\label{fig:quality}
\end{figure}

Figure~\ref{fig:quality}
summarizes the parameters characterizing the light profile fit quality 
and the galaxy sample for the ENEARm and ENEARc, showing no significant
differences between either samples.  
We found that the quality parameter Q $\leq$ 2 for  more than 80\% of
ENEARm  galaxies and more than 70\% for ENEARc galaxies (panel a).
Furthermore, galaxies belonging to ENEARc
tend to have slightly smaller  $D/B$ (panel b), and as expected smaller
$r_{max}$/$r_e$ (panel c) and smaller $S/N$ (panel h). 
Consequently, the ENEARc fit qualities are
slightly worse than in the case of ENEARm, since, in general, galaxies
with smaller apparent sizes are more affected by seeing (panel d) and
require larger extrapolations (panels c and i). However, this
is not so evident in Figure ~\ref{fig:quality}
 because the majority of the ENEARc
galaxies (201 out of 335) belongs to ENEARm. The ENEARm is more uniform 
with slightly better
fits as shown in panel (e). The calculation of total
magnitudes requires extrapolations
smaller than 10\% in 70\% (60\%) of the cases for ENEARm (ENEARc)
galaxies. Analogous plots may be seen in Saglia \etal
(1997) for the EFAR sample of rather more distant clusters to $cz \sim $
18,000 \kms.  The differences between ENEAR and EFAR
are apparent by comparing panels (a) and (e) with their equivalents in 
Saglia \etal (1997).  Our sample contains brighter galaxies and we have
a larger percentage of objects with the best quality fits.  Also the
differences with panels (b), (d) and (g) underline the difficulties 
of selecting true ellipticals of small apparent sizes.
The seeing tends to bias the results for more distant samples as
there is a preference in selecting higher surface brightness 
objects, when compared to the nearby samples.

\subsection {Structural Parameters}

\subsubsection{Homogenization}
\label{homog}

When constructing the homogeneous data set
of structural parameters used in the derivation of scaling relations,
any remaining systematic shifts in the parameters
were minimized using multiple observations of
the same galaxies.  We use the measurements obtained from images taken
at ESO with setups 4 and 5 (see Table \ref{tab:setups}) as our fiducial
system. These setups were chosen because they have the largest number of
repeated observations using the same telescope, similar detectors,
present the largest field of view and have the best resolution.

To determine the ``fiducial'' system we corrected our photometric
parameters using the mean difference 
$ \Delta y_{i} = \epsilon_{i}^2 \sum\limits_{j\neq i} \sum\limits_{k \in {i,j}} \frac{y_{ik}-y_{jk}}{\Delta y_{ik}^2 +\Delta y_{jk}^2} $ 
between the measurements
of run $i$ with all the other runs $j \ne i$ for galaxies in run $j$ in common
with those in run $i$. This offset is computed weighting by the variance which we
take as being the estimated errors 
$\epsilon_{i} = \left(\sum\limits_{j\neq i} \sum\limits_{k \in {i,j}} \frac{1}{\Delta y_{ik}^2 +\Delta y_{jk}^2} \right) ^{-1/2}$
 in each measurement. Here $k$ runs over the galaxies
in common to runs $i$ and $j$, and $y_{ik}$ corresponds to the 
measurement of either $D_n$ or $r_e$ or \mue for galaxy $k$ 
in run $i$ and $\epsilon_{ik}$ is the estimated error. 

We determined the most significant offset by finding the run with the
maximum value $\Delta y_i/ \sigma_i$, where $\sigma_i$ is the standard
error in the mean of run $i$, and iterated towards a common
zero-point by subtracting this offset from the measurements of run
$i$.  We finished the process when the most significant offset was
$\Delta y_i/ \sigma_i <2$.  After three iterations the systematic
offsets required to create a homogeneous fiducial data set were
determined. In general, we found good agreement between the
photometric parameters measured from repeated observations, so that
the corrections required to bring them into a common system were
relatively small: $\Delta log\,D_n \lsim 0.010$, 
$\Delta $ \mue $\lsim 0.08$ mag arcsec$^{-2}$, 
$\Delta log\,r_e \lsim 0.03$ and $\Delta$ m$_{R_C} \lsim 0.03$.
In this process we
discarded galaxies which exhibited peculiarities in their profiles as
indicated by the comments in Table~\ref{tab:data} (explained below).

After defining this standard system, the structural parameters derived from
observations obtained at MDM, FLWO and CTIO were also transformed into it.  For
runs with a significant number of galaxies in common with our
reference system, the measured values were directly compared,
while other runs were corrected using the calibrated measurements for the
same telescope. The
corrections required for all data sets were small zero-point shifts,
typically: $\Delta log\,D_n \lsim 0.003$, $\Delta $ \mue $ \lsim
0.04$ mag arcsec$^{-2}$, and $\Delta \log \,r_e \lsim 0.010$. These
corrections are
comparable to those found when defining the reference system.

\subsubsection{Results}

\begin{figure}
\vspace{5cm}
\centering
\mbox{\psfig{figure=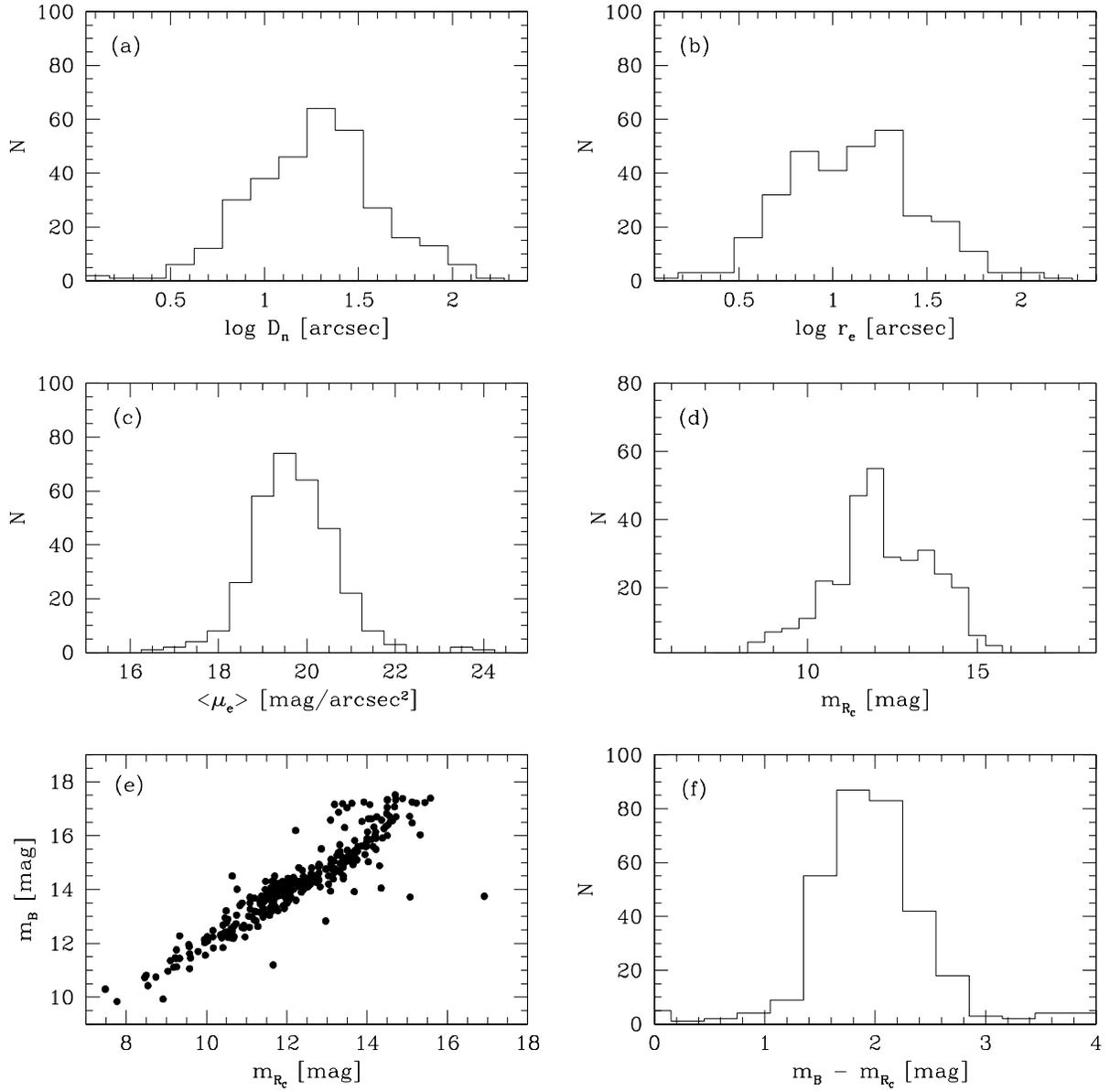,height=13truecm,bbllx=2truecm,bblly=5truecm,bburx=19truecm,bbury=20truecm }}
\caption{The same plots of Figure~\ref{fig:histphotenearf} for
galaxies in clusters (ENEARc sample)}.
\label{fig:histphotenearcf}
\end{figure}

The structural parameters derived for the ENEARm and ENEARc
samples are presented in
Figures~\ref{fig:histphotenearf} and \ref{fig:histphotenearcf},
where we show: in panel
(a) the distribution of $log$ \dn ($arcsec$); in panel (b) the distribution
of $log$ \re (also in arcsec); in panel (c) the surface brightness
distribution; in panel (d) the distribution of the total
R$_C$- magnitudes; in panel (e) the relation between B magnitudes (from
the literature), and our R$_C$ magnitudes;
and finally in panel (f) the (B--R$_C$) colors.

Comparing these figures clearly shows the differences
between the ENEARm and ENEARc samples, even though there is considerable
overlap between them. As seen in panel (d) the ENEARm sample has a reasonably
well-defined limiting magnitude at
R$_C$ = 13.0 which agrees well with the magnitude-limit
adopted in the photographic B-selected sample where errors in the magnitudes
can be as large as 0.5~mag (Alonso \etal
1993). While there is a linear relation between the B and
R$_C$ magnitudes, the scatter is large especially at the faint
end. This is also seen in panel (f) were the mean colors are near 1.9.  
These differences with the adopted color values of 1.48 to calibrate the 
photometry are mainly due to the fact that B magnitudes are photographic and
the R$_C$ magnitudes are obtained by the component luminosities 
 of the fit decomposition.  Colors redder than 2 indicate unreliable 
photographic B magnitudes or  unreliable fits
with higher extrapolations, the main source of uncertainties in $m_{R_C}$.
The angular
size of galaxies in the ENEARm (panels a and b), as measured either by \dn or
\re, is large, with a median of $\sim 23$~arcsec and $\sim 17$~arcsec,
respectively. As discussed above, this means that the structural
parameters for
the ENEARm sample are rather insensitive to seeing, in contrast
to the cluster ENEARc sample shown in Figure~\ref{fig:histphotenearcf}
that has median \dn and \re
values respectively of about 18 and 15 arcsec, 
with an extended tail towards smaller values.
As the ENEARc is not a strictly magnitude-limited
sample, galaxies can cover a wide range of apparent sizes, and thus
present a wider distribution of structural parameters, specially in \re
(panel b).
Finally,  the distributions of the
mean effective surface brightnesses are similar, peaking near $\sim
19.5$~\bri in R$_C$ but with a wider distribution in the ENEARc sample.

\subsubsection{Internal and External Comparisons: Errors}
\label{errors}

\begin{figure}
\vspace{5cm}
\centering
\mbox{\psfig{figure=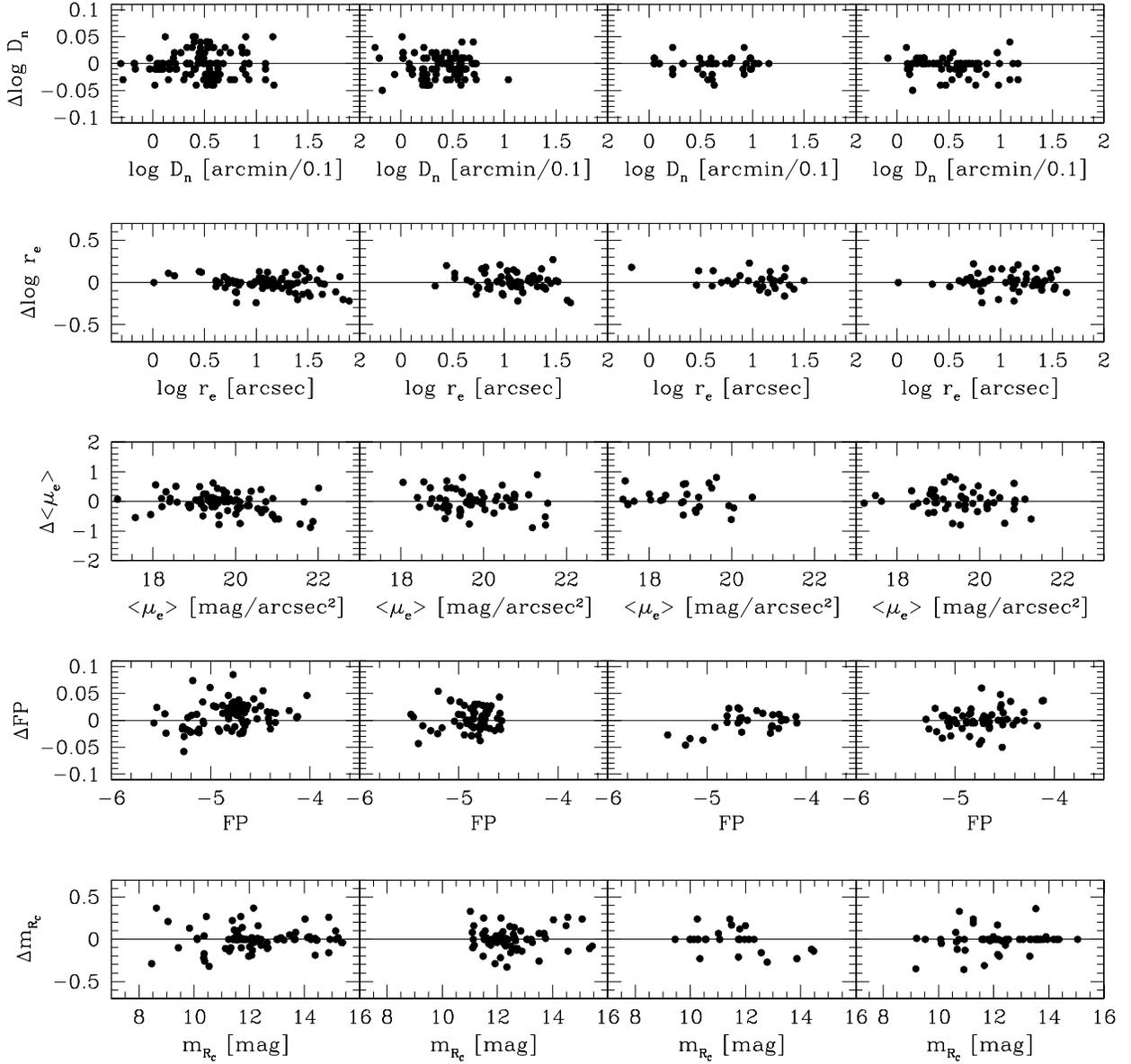,height=13truecm,bbllx=2truecm,bblly=5truecm,bburx=19truecm,bbury=20truecm }}
\caption{
Internal comparisons of $log\,D_n$, $log \, r_e$, $<\mu_e>$, FP, and
m$_{R_C}$ derived from data observed at ESO, MDM, FLWO and CTIO (panels from left
to right, respectively).}
\label{fig:comparphint}
\end{figure}

\begin{figure}
\vspace{5cm}
\centering

\mbox{\psfig{figure=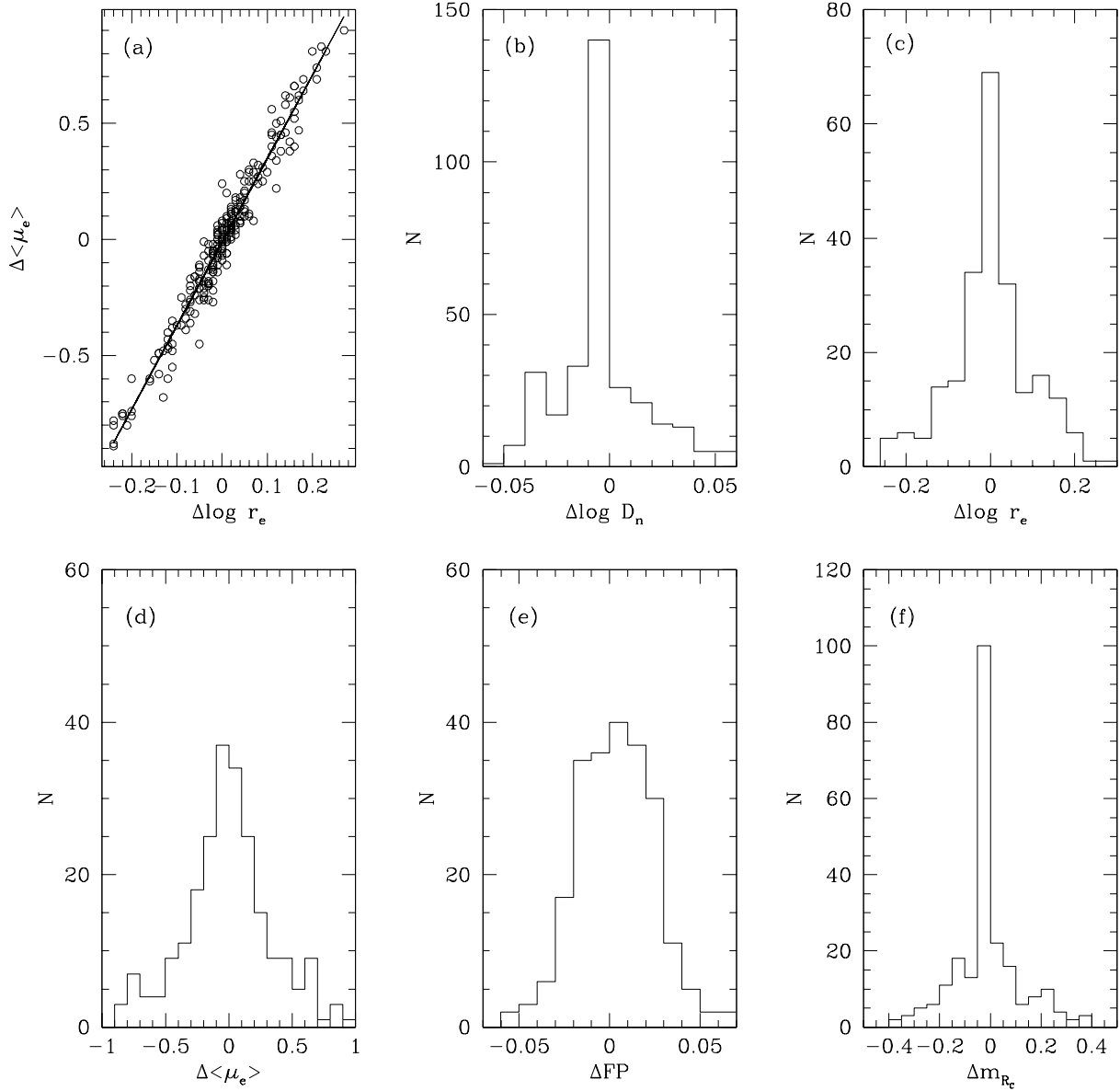,height=13truecm,bbllx=2truecm,bblly=5truecm,bburx=19truecm,bbury=20truecm }}
\caption{The distribution of the differences between parameters derived from 
our internal comparison:  (Panel a) shows the correlation of the differences
in $log \, r_e$ and $<\mu_e>$ and the solid line is the expected relation
$log \, r_e$ = 0.27 $<\mu_e>$.  The histograms of the differences are shown in
 (Panel b) $log\,D_n$, (Panel c) $log \, r_e$, (Panel d) $<\mu_e>$, (Panel e)
FP and (Panel f) $m_{R_C}$.}
\label{fig:histodiff}
\end{figure}

 In Figure~\ref{fig:comparphint} 
we compare the structural parameters of our repeated observations after
applying the zero-point corrections. Again, we performed the differences
using the convention  "older minus newer" measurements, as explained in
Section~\ref{surf}.   We show, from top
to bottom, the differences between the calibrated measurements of \ldn,
\lre, \mue, FP and the total magnitude (m$_{R_C}$)
observed at different sites (see figure caption).  In the Figures, 
the dispersion in \ldn is smaller than for \lre and \mue.
\dn is obtained by a simple interpolation while \re and \mue
result from fits where extrapolations are important.  In the case of 
small CCDs, the extrapolations are the main source of uncertainties.
The largest scatter in the photometric parameters occurs with FLWO data
but no single telescope set has a much larger uncertainties than the others.
Figure~\ref{fig:histodiff} shows the distribution of the differences of 
all the compared parameters.   The correlation of the differences
in $log \, r_e$ and $<\mu_e>$ is shown in panel (a). The solid line 
shows the relation $log \, r_e$ = 0.27 $<\mu_e>$ in good agreement
with JFK results.   The histograms 
of the differences are shown in (Panel b)  $log\,D_n$, (Panel c) $log \, r_e$, 
(Panel d) $<\mu_e>$, (Panel e) FP and (Panel f) $m_{R_C}$.  The distribution
of differences in FP is broader than the distribution of the differences in
$log\,D_n$.

Table~\ref{tab:compglobstat} summarizes the results of
these comparisons which
gives: in column (1) the site; in columns (2) the number of repeated
measurements $N_{D_n}$ in that site or in common with our standard system;
in column (3) the mean offset and error of \ldn; and in column (4) its
scatter. In the remaining columns the same information, including the
number of repeated measurements $N_{FP}$, is given for
\lre, \mue, FP and $m_{R_C}$.  These results are comparable to those 
internal estimates by JFK.

We transformed all the photometric data into the fiducial system and
combined the individual measurements weighting by the errors
in the cases of multiple observations.  For each value of $D_n$ the 
error was computed taking into account the 
uncertainties associated with the quality of the fits,  and the rms 
scatter $\sigma_{D_n}$ measured from multiple observations of galaxies
obtained using the same telescope. These errors include the contribution 
from the photometric calibration 
because they were scaled from the internal comparisons.  Its
effect was also estimated from Monte-Carlo simulations
where a number of light profile shapes covering the range of
magnitudes and D/B ratios observed in the ENEAR sample were generated.
For each of these galaxy profiles, 100 simulations were created
by shifting the photometric zero-point by an
offset drawn from a Gaussian deviate with a dispersion equal to the
estimated zero-point error of the calibration.  For each of these
simulated profiles the photometric parameters were calculated and their
mean and scatter were computed. This was done for a range of $rms$
values up to 0.05~mag, the value we adopted for a
photometric night. We found that the errors in \dn depend on the profile 
shape, and the
zero-point errors lead to uncertainties which are comparable to
those estimated for high quality profiles (see Bernardi 1999, for details). 
Therefore, we prefer to estimate the final errors for \dn taking into 
account the quality of the fit and the scatter 
of the internal comparisons. 

\begin{figure}
\vspace{5cm}
\centering
\mbox{\psfig{figure=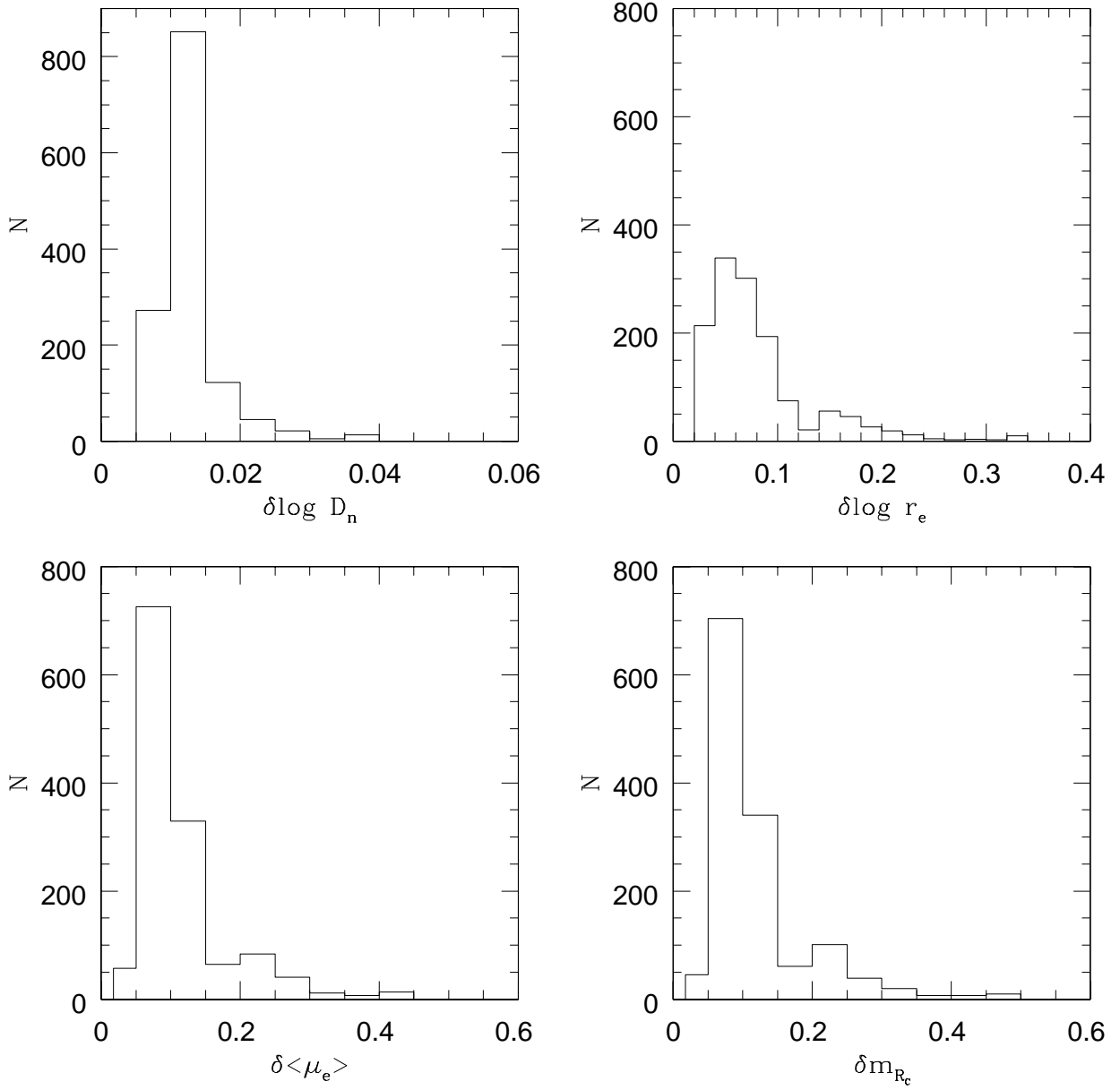,height=13truecm,bbllx=2truecm,bblly=5truecm,bburx=19truecm,bbury=20truecm }}
\caption{The distribution of the errors of the following parameters,
obtained as explained in Section~\ref{results}:
(Panel a) $log\,D_n$, (Panel b) $log \, r_e$, (Panel c) $<\mu_e>$, and
(Panel d) $m_{R_C}$.}
\label{fig:pherr}
\end{figure}

The same procedure was applied to derive errors for the other
parameters, but in these cases the main contribution is 
the uncertainty associated with the quality of the fits.
The distribution of all error parameters of interest
(\ldn, \lre, \mue, and $m_{R_C}$), is shown in Figure~\ref{fig:pherr}
for galaxies in the ENEARm sample.
The median errors are 0.011 dex for $\log{D_n}$, 0.064 dex
for $\log{r_e}$, 0.086~\bri for \mue and 0.09 mag for $m_{R_C}$. 

\begin{figure}
\vspace{5cm}
\centering
\mbox{\psfig{figure=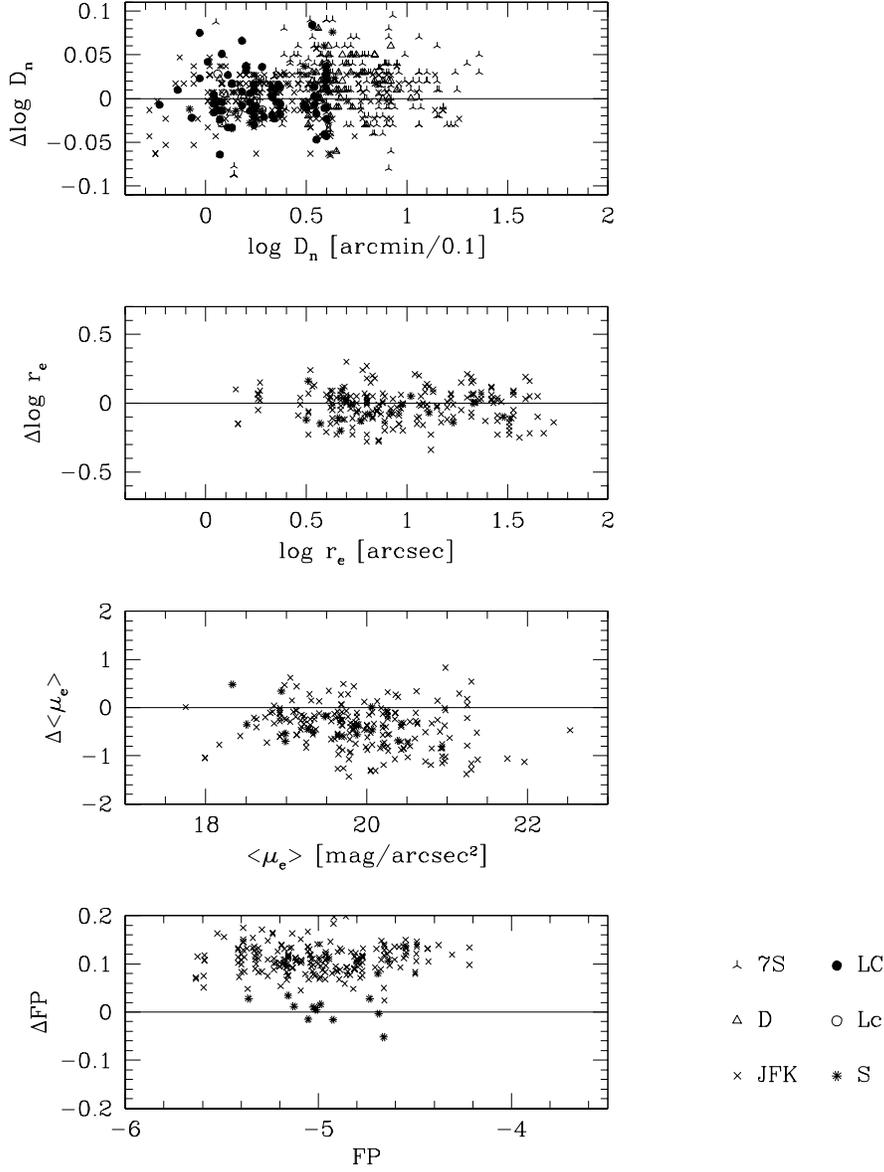,height=13truecm,bbllx=2truecm,bblly=5truecm,bburx=19truecm,bbury=20truecm }}
\caption{ The overall external comparison of $log{D_n}$, $log \, r_e$,
$<\mu_e>$, and FP derived from our data.
$\Delta$ means ``ours-literature'' measurements. The literature
sources are: LC, Lucey \& Carter (1988); 7S, Faber \etal (1989); D, Dressler 
(1987) and Dressler, Faber \& Burstein (1991);
JFK, J\o rgensen, Franx, \&  Kj\ae rgaard (1995); Lc, Lucey
\etal (1997); and S, Smith \etal (1997).}
\label{fig:comparphlit}
\end{figure}

The accuracy of our measurements was tested by comparing our 
structural parameters with
those of other authors. Altogether there are 354
galaxies in our sample in common with  Dressler (1987); 
Lucey \& Carter (1988);
 Faber \etal (1989); Dressler, Faber \& Burstein (1991); JFK;  
Lucey \etal (1997); and Smith et al. (1997).
The largest overlaps are with Faber et al. (1989), with whom we have
293 galaxies in common, and JFK with 232 galaxies.  These
comparisons are shown in Figure~\ref{fig:comparphlit}.
Because the errors in $r_e$ and \mue
correlate strongly (J\o rgensen \etal 1996) and they are sensitive to
the fitting procedure adopted by different authors,
we also present the comparison for the photometric components of the
FP, that is \lre -- 0.30\mue. Not all parameters are available in the 
literature, especially in the case of those derived from profile fits. 

The results of these comparisons are presented in
Table~\ref{tab:compglobext} listing the literature source; the number of 
galaxies used in the \dn comparison; the differences, error and scatter 
for \dn; the number of galaxies used in the comparisons of \re,
\mue and FP and the differences, error and scatter, respectively.  
In the \dn comparison, all the values are defined in the B-band isophotal 
level and the results are consistent with this assumption.  The larger
offset in the comparison of \ldn with 7S and Dressler (1987) values was
also found by JFK.  The differences observed with JFK for the FP
parameters, specially an offset of about 0.4 in \mue, reflect the color
term introduced between their Gunn r data and our Kron-Cousins photometry. 
Taking into account that there is a mean difference in magnitudes of r-R$_C$ 
of about 0.354 mag (J\o rgensen 1994) and in $log\,r_e $ of -0.014 
(Table~\ref{tab:compglobext}), it is straightforward to find a difference
in \mue of 0.424, in agreement with our results. A similar 
offset was also found by Smith et al. 
(1997) in their comparisons with JFK.  With them we are in the same system
and all the galaxies we have in common are in the central parts of 
clusters.  The observed difference in the \mue comparison is due mainly to
strong light contamination from nearby galaxies, giving a difficult sky 
subtraction and more uncertain fit results.  Once again, \dn is obtained
in the inner parts and the light contamination is not so important.
The observed scatter in \dn in all cases is nearly the same, $\sim$ 0.025,
consistent with our internal error estimates ($\sim$ 0.023), if we assume that
our errors and those of the other authors are of the same amplitude.
The scatter for the other parameters in the comparisons with JFK
is consistent with their results for external comparisons and also with 
Saglia et al. (1997) results.

\section{The ENEAR Photometric Catalogue}
\label{catalog}

The final catalogue listing structural parameters for 1332 galaxies is
presented in Table~\ref{tab:data}. The Table comprises 1104 galaxies in
the ENEARm sample; 335 in ENEARc, of which 201 are also contained in
ENEARm; and 94 galaxies with later morphological
types, as explained in Section~\ref{sample}. The Table gives in
column (1) the galaxy identifications from NGC, IC, ESO, MCG, UGC
catalogs.  For galaxies in clusters not in these 
catalogs, we use D from Dressler (1980), RH and RK 
(Hydra cluster and Klemola 27, respectively from Richter 1989), 
WA and WS (Abell 3574--Klemola 27 and 
S753, respectively from Willmer \etal 1991), J from JFK, 
ZH from
Zwicky \& Humason (1964) and L from Lucey \etal (1991).  
The Table also lists in columns (2)-(3) the 2000 equatorial coordinates;
in columns (4) the morphological type $T$
(Lauberts \& Valentijn 1989); (5)  B(0) magnitude and
(6) radial velocity, all from the literature; in column (7) the number of
our observations; in columns (8)-(9) the total R$_C$-band magnitude and its error;
 in columns (10)-(15) the same information for \ldn ($D_n$ in arcmin/0.1);
\lre ($r_e$ in arcsec) and \mue (in mag/arcsec$^2$), respectively; in
column (16) the D/B ratio; in column (17) the FWHM of the
point-spread function (in arcsec) determined from the fit; in column 
(18) are notes describing
features observed in the galaxy light profile which may
affect the determination of the photometric parameters;
and in column (19) the galaxies previously observed by other
authors (indicated with an asterisk). 
The column containing the notes
also identifies objects that may have to be excluded in analyses
requiring reliable photometric parameters, such as the
derivation of galaxy distances, one of our primary goals.

\section{Summary}

We have presented structural parameters derived from the surface
photometry for 1332 early--type galaxies as part of the ENEAR project.
These galaxies have been used in previous papers of this series to probe the
peculiar velocity field in a
volume within 7000 \kms. The present sample represents a wide-angle
photometric survey of early--type galaxies brighter than \mb = 14.5
mag, which extends the 7S sample both in morphological types and
depth.  It also complements the recently completed TF surveys of spiral 
galaxies
(Mathewson \etal 1992, 1996, da Costa \etal 1996, Haynes \etal 1999
a,b) for studies of the peculiar velocity field in the nearby Universe.

The surface brightness profiles have been obtained from circular
apertures, and have been fit by a seeing--convolved two--component bulge--disk model 
to derive structural parameters such as \dn, \re and \mue.
We evaluate the quality of the data comparing the profiles
and structural parameters obtained from multiple observations of individual
galaxies. These multiple observations were also used to bring the
photometric parameters into a common, homogeneous fiducial system.
The corrections were relatively small: $\Delta log\,D_n \lsim 0.003$, 
$\Delta $ \mue $\lsim 0.04$ mag arcsec$^{-2}$, and
$\Delta log\,r_e \lsim 0.010$.
The errors in the structural parameters estimated from the scatter of
internal comparisons  are: 0.011~dex in \ldn, 0.064~dex in \lre,
0.086~\bri in \mue  and 0.09 for $m_{R_C}$.
The comparison of our photometric data for galaxies  in common with
other authors shows good agreement and confirms that our
internal error estimates are fairly robust.

From the profile fitting we found that about 12 \% of the galaxies are well
represented by a
pure $r^{1/4}$ law while 87\% are best fit by a two component model.
In the original morphological classification of the galaxies there are
 about 26\% of ellipticals.
In general we find that 60\% of ENEARm galaxy profiles are of high
quality, while for galaxies in clusters (ENEARc) this number decreases
to about 47\%.   The derived photometric
parameters have been used, in conjunction with the spectroscopic data (Wegner 
et al. 2003), in previous papers of this series  to derive reliable 
distances and 
map the peculiar velocity field in the nearby Universe.

\noindent {\bf Acknowledgements}

We would like to thank the anonymous referee to help us to improve the
text with constructive comments.  
MVA would like to thank Wolfram Freudling, 
Inger J\o rgensen, David Latham, Alejandra Milone,
Reynier Peletier, and Ivo Busko for their contributions during
different phases of this project.  MVA and MB thank
Roberto Saglia for providing and assisting us with the program to
obtain seeing corrected decompositions of the profiles.
MVA would like to thank the
hospitality of the Harvard--Smithsonian Center for Astrophysics, the
ESO visitor program and ON where much of the work was carried
out. We wish to thank the support of the CNPq--NSF bilateral program
(MVA, LNdC), a research fellowship from CNPq (PSSP) and CLAF (MVA, PSSP and
MAGM) for financial support to the project. MB also thanks the Sternwarte 
M\"unchen, the Technische
Universit\"at M\"unchen, ESO Studentship program, and MPA Garching for
their financial support during different phases of this research.
 GW acknowledges support from the following over the
course of this project: Dartmouth College, 
the Alexander von Humboldt-Stiftung for a year's
stay at the Ruhr Universit\"at in Bochum, and ESO for supporting trips to
Garching.  MVA also acknowledges financial
support from the SECYT and CONICET (Argentina). CNAW acknowledges
partial support from CNPq grants 301364/86-9, 453488/96-0, and NSF
AST-9529028 and NSF AST-0071198

{}


\normalsize
\begin{table}
\begin{center}
\caption{Observing Runs for Photometry}
\vspace*{10truemm}
\tiny
\begin{tabular}{lcrr}
\hline \hline
\\[-4mm]
\label{tab:runs}
Run & Date & N$_{tp}$ & Setup \\
 (1) &  (2) & (3) & (4) \\
\\[-4mm]
\hline
\\
CTIO-701 & Nov 87& 5/3 & 9\\
FLWO-201 & Dec 88& 15/0 & 7\\
FLWO-202 & Apr 89& 11/8 & 7\\
FLWO-203 & Sep 89& 9/6 & 7\\
ESO-601 & Nov 89 & 4/4 & 1\\
CTIO-702 & sep 90& 2/0 & 9\\
FLWO-204 & Nov 91& 4/0 & 8\\
ESO-602 & Sep 92 & 4/0 & 1\\
FLWO-205 & Oct 92& 6/2 & 8\\
MDM-551 & Jan 93 & 4/0 &10\\
FLWO-206 & Mar 93& 4/3 & 8\\
ESO-603 & Jul 93 & 5/3+1 & 2\\
ESO-604 & Nov 93 & 4/2+2 & 2\\
ESO-605 & May 94 & 7/5 & 2\\
MDM-552 & Mar 95 & 7/5+1 &11\\
ESO-606 & Aug 95 & 2/2 & 3\\
MDM-553 & Nov 95 & 7/6+1 &12\\
ESO-611 & Dec 95 & 16/15 & 6\\
MDM-555 & May 96 & 3/3 &12\\
ESO-607 & Oct 96 & 3/0 & 4\\
MDM-554 & Nov 96 & 8/0 &12\\
ESO-608 & Feb 97 & 3/3 & 4\\
MDM-556 & Feb 97 & 6/3 &12\\
ESO-609 & Apr 97 & 4/4 & 5\\
MDM-557 & Jun 97 & 4/3 &12\\
ESO-610 & Nov 97 & 4/1 & 5\\
MDM-558 & Nov 97 & 4/0 &12\\
ESO-613 & Mar 98 & 3/3 & 5 \\
MDM-559 & May 98 & 4/3 &12 \\
MDM-560 & Nov 98 & 1/1 &12 \\
CTIO-703 & Feb 99& 7/6+1 &10\\
CTIO-704 & Sep 99& 7/0 &10\\
\\
\hline
\end{tabular}
\end{center}
\footnotesize
Notes: In column (3) is reported 
the number of the total/photometric nights for the 
corresponding run.  This column also
includes the number of partially photometric nights, preceded by a
plus sign. Information about
the setup indicated in column (4) are given in Table~\ref{tab:setups}
\end{table}

\small
\begin{deluxetable}{rlrrlrrrr}
\tabletypesize{\scriptsize}
\tablenum{2}
\label{tab:setups}
\tablecaption{Observing Setups}
\tablewidth{0pt}
\tablehead{
\colhead{Setup} & \colhead{Telescope} & \colhead{$N_{m}$} & \colhead{$N_{r}$} & \colhead{Detector} & \colhead{Field of View} & \colhead{Scale} & \colhead{Gain} & \colhead{RON} \\
      &           &  & &        & arcmin$\times$ arcmin & arcsec/pixel & e$^-$/ADU &  [e$^-$]}
\startdata
1 & DK 1.54m & 36 & 8 & RCA 5264-7-3 & $4 .0 \times 2 .5$& 0.47 & 20 & 15 \nl
2 & DK 1.54m & 414 & 61 & Tek \#28 & $6 .5 \times 6 .5$ & 0.38 &  3.5  & 8.0 \nl
3 & DK 1.54m & 153 & 44  & CCD \#17 & $8 .5 \times 8 .5$& 0.51 & 2.0 & 3.7 \nl
4 & DK 1.54m & 96 & 17 & LORAL/LESSER W11-4& $13 .3 \times 13 .3$  &0.39 &  1.31 &  7.2 \nl
5 & DK 1.54m & 304 & 54 & LORAL/LESSER C1W7& $13 .3 \times 13 .3$ &0.39 &  1.31 &  7.2 \nl
6 &  Dutch 0.9m & 50 & 21 & Tek                         & $3.8 \times 3.8$ & 0.44 & 3.56 & 8.0 \nl
  &     & 1053 & 205  &         &       & & \nl
  &     &      &      &         &       & & \nl
7 &  FLWO 0.61m & 85 & 8 & Tek & $5 .5 \times 5 .5$   &  0.65 &  3.8  &   12 \nl
8 &  FLWO 1.3m & 228 & 45 &Tek & $11 .2 \times 11 .2$&  0.65  &  2.5  &   13 \nl
  &      & 313 & 53  &            &       & & \nl
  &     &      &      &         &       & & \nl
9 &  CTIO 0.9m & 47 & 26 & RCA \# 5 & $4 .2 \times 2 .6$  &  0.49 &  6.5  &  -- \nl
10&  CTIO 0.9m & 247& 42 & Tek2K & $13 .5 \times 13 .5$    & 0.396 &  3.2  & 4.0 \nl
  &      & 294 & 68  &         &       & & \nl
  &     &      &      &         &       & & \nl
11 &  MDM 1.3m & 48 & 4 & Wilbur LORAL binned 2$\times$2 & $10 .5 \times 10 .5$ & 0.63 & 2.25 & 4.73 \nl
12 &  MDM 1.3m & 413  & 26 & Nellie STIS& $15 .0 \times 15 .0$ & 0.44 & 2.94 & 4.38 \nl
  &      & 461  & 30      &         &   & &      \nl
  &     &      &      &         &       & & \nl
Total &  & 2121 & 356  &         &   & &     \nl
\enddata
\footnotesize 
\vskip 0.5cm
Notes: In columns (3) and (4) are listed the total number of images observed 
in the R$_C$ band ($N_m$) in the different 
setups and the number of repeated images ($N_r$) for that setup,
which are used as calibrators to homogenize our observations.
\end{deluxetable}

\clearpage

\normalsize
\begin{table}
\begin{center}
\tablenum{3}
\caption{Internal Comparisons of the Light Profiles}
\vspace*{10truemm}
\begin{tabular}{lrcc}
\hline \hline
\\[-4mm]
\label{tab:compstatint}
Comparison & N$_c$ & $ \Delta \mu$ & $\sigma_{\mu}$\\
 &  & mag/arcsec$^2$ & mag/arcsec$^2$\\
\\[-4mm]
\hline
\\
same night  & 172 & 0.004 $\pm$ 0.003 & 0.042 \\
same setup  & 53  & 0.001 $\pm$ 0.007 & 0.052 \\
different setups & 114 & 0.002 $\pm$ 0.009 & 0.094 \\
\hline
\end{tabular}
\end{center}
\end{table}

\clearpage

\tiny
\begin{table}
\tablenum{4}
\label{tab:compglobstat}
\tabcolsep=0.15cm
\caption {Internal Comparisons of the Structural Parameters}
\vspace*{10truemm}
\tiny
\begin{tabular}{lcccccccccccc}
\hline \hline
\\[-4mm]
Site & $N_{D_n}$ & $\Delta log\,D_n$ &
$\sigma_{log\,D_n}$& $ N_{FP}$ & $\Delta log \,r_e$ &
$\sigma_{log \,r_e}$ &$ \Delta $ \mue &
$\sigma_{<\mu_e>}$ & $ \Delta FP$ & $\sigma_{FP}$&
 $\Delta m_{R_C}$ & $\sigma_{m_{R_C}}$ \\
  &  & arcmin/0.1 & & & arcsec &  & mag/arcsec$^2$ & & & & mag & \\
\\[-4mm]
\hline
\\
ESO & 118 & -0.001$\pm 0.002 $ & 0.023 & 82 &
-0.013$\pm 0.010$ & 0.089 & -0.074$\pm 0.036$ & 0.327 &
0.010$\pm 0.003$ &0.024 &-0.001$\pm 0.014$ & 0.129 \\
MDM & 88 & -0.005$\pm 0.002$ & 0.020 & 63 &
 0.006$\pm 0.013$ & 0.100 & 0.016$\pm 0.046 $ & 0.365 &
0.001$\pm 0.003$ & 0.023 & 0.001$\pm 0.017$ & 0.132\\
FLWO & 43 & -0.002$\pm 0.002$ & 0.014 & 29 &
0.021$\pm 0.017$ & 0.092 & 0.083$\pm 0.064$ &0.346 &
-0.004$\pm 0.003$ & 0.018 & -0.019$\pm 0.025$ & 0.133 \\
CTIO & 64 &-0.005$\pm 0.002$ & 0.017 & 55 &
0.008$\pm 0.013$ & 0.094 & 0.025$\pm0.046$ & 0.343 &
0.001$\pm0.003$ & 0.021 & -0.012$\pm 0.017$ & 0.126  \\
\hline
\end{tabular}
\end{table}

\clearpage
\tiny
\begin{table}
\tablenum{5}
\label{tab:compglobext}
\tabcolsep=0.15cm
\caption {External Comparisons of the Structural Parameters.} 
\vspace*{10truemm}
\tiny
\begin{tabular}{lccccccccccc}
\hline \hline
\\[-4mm]
Sources & N$_{D_n}$ &$\Delta log\,D_n$ &
$\sigma_{log\,D_n}$& N$_{FP}$ &  $ \Delta log \,r_e$ &
$\sigma_{log \,r_e}$ &$ \Delta $ \mue &
$\sigma_{<\mu_e>}$ & $ \Delta FP$ & $\sigma_{FP}$ \\
 &  & arcmin/0.1 & & & arcsec & & mag/arcsec$^2$ & & & \\
\hline
LC  & 84 & 0.003 $\pm 0.003$ & 0.027 & ... & ... & ... & ... & ... & ... & ... \\
7S      & 293 & 0.013$\pm 0.002$ & 0.034 & ... & ... & ... & ... & ... & ... & ... \\
D       & 54  & 0.016$\pm 0.004$ & 0.027 & ... & ... & ... & ... & ... & ... & ... \\
JFK & 232 & 0.002$\pm 0.002$ & 0.023 &
210 & -0.014$\pm 0.008$&  0.116& -0.410$\pm 0.030$ & 0.431& 0.109$\pm 0.003$  & 0.039 \\
Lc  & 13 & -0.004$\pm 0.004$ & 0.015 & ... & ... & ... & ... & ... & ... & ... \\
S   & 15 &  0.005$\pm 0.009$ & 0.034 &
13 & -0.074$\pm 0.026$& 0.095 & -0.306$\pm 0.103$ & 0.372 &0.018$\pm 0.011$ & 0.041\\
\hline
\end{tabular}
\\[3mm]
\footnotesize
Notes: All differences are ``our measurement'' -
``literature measurement''.\\
The references are 
LC: Lucey \& Carter (1988);
7S: Faber \etal (1989);
D: Dressler (1987) and Dressler, Faber \& Burstein (1991);
JFK: J\o rgensen, Franx, \&  Kj\ae rgaard (1995);
Lc: Lucey \etal (1997);
S: Smith et al. (1997)
\end{table}


\clearpage

\clearpage

\clearpage

\clearpage

\clearpage

\clearpage

\clearpage

\clearpage

\clearpage

\clearpage

\clearpage

\clearpage

\tabcolsep=0.10cm
\textheight=10in
\textwidth=7.5in
\oddsidemargin  .30in
\evensidemargin .07in
\topmargin=-0.2in
\rotate{
\begin{planotable}{lccrrrrrrcrrrccrclr}
\small
\tablewidth{0pt}
\tablenum{6}
\tablecaption{The Photometric ENEAR Catalog}
\tablehead{
Name & $\alpha$& $\delta$&      T & m$_{B}$& $cz_{hel}$ & N$_{obs}$ &$m_{R_C}$& $\epsilon_{m_{R_C}}$ & $log\, D_n$ &      $\epsilon_{D_n}$& $log \, r_e$ &     $\epsilon_{r_e}$& $<{\mu_e}>$ & $\epsilon_{<\mu_e>}$   & D/B & FWHM & Notes & Lit\\
 & (2000) & (2000) & & mag & km s$^{-1}$ & &mag& & arcmin/0.1 & & arcsec & & mag/arcsec$^2$ & & & arcsec&      & \\
(1) & (2) & (3) & (4) & (5) & (6) & (7) & (8)      & (9) & (10) & (11) &     (12) & (13) & (14) & (15) & (16) & (17) & (18) & (19) }
\startdata
NGC7805       & 00:01:27.1 & +31:26:02     & -2 & 14.30 &  4948 &  1 & 12.75 & 0.05 & 0.490 &  0.010 &  0.71 &  0.04 & 18.25 &  0.05 &  0.17 & 0.64 &  1           &    \nl
NGC7810       & 00:02:19.3 & +12:58:16     & -2 & 14.30 &  5532 &  1 & 12.26 & 0.09 & 0.520 &  0.012 &  1.10 &  0.07 & 19.66 &  0.09 &  0.24 & 1.35 &  1           &    \nl
NGC7832       & 00:06:28.4 & -03:42:58     & -3 & 13.50 &  6204 &  1 & 11.72 & 0.11 & 0.570 &  0.012 &  1.25 &  0.08 & 19.95 &  0.11 &  0.05 & 1.94 &  2           &    \nl
UGC00061      & 00:07:23.8 & +47:02:26     & -2 & 14.30 &  5277 &  1 & 12.06 & 0.30 & 0.580 &  0.024 &  1.06 &  0.21 & 19.28 &  0.28 &  0.00 & 1.77 &  ...         &    \nl
NGC0043       & 00:13:00.8 & +30:54:55     & -2 & 13.90 &  4785 &  1 & 11.72 & 0.25 &  0.520 &  0.020 &  1.37 &  0.17 & 20.50 &  0.23 &  1.56 & 1.15 &  7           &    \nl
UGC00130      & 00:13:56.9 & +30:52:58     & -7 & 14.20 &  4735 &  1 & 13.10 & 0.05 & 0.420 &  0.010 &  0.72 &  0.04 & 18.66 &  0.05 &  0.30 & 0.71 &  ...         &    \nl
\enddata
\tablecomments{
 The number in column (18) flags the following causes for features observed in the
image and/or spectrum of a galaxy:                            
(1) strong contamination  by other galaxies along the line of sight, or
interacting galaxies;                         
(2) strong contamination by bright stars along the line of sight;              
(3) crowded background;                                         
(4) presence of spiral arms or shells;                                     
(5) presence of a bar; 
(6) presence of dust lanes;                 
(7) high D/B ratio, edge-on galaxy;                               
(8) evidence of star formation; 
(9) peculiar shape, peculiar nucleus, presence of spikes;
(10) faint galaxy; 
(11) $D_n$ available in the literature but uncertain or image problems: large masked region, saturation, large galaxy compared to the field-of-view,
faint parts near the CCD limits; 
(12) presence of a halo or ring in the galaxy;
(13) dwarf galaxy;
(14) galaxy observed with poor seeing or elongated PSF;
(15) peripheral cluster member.}
\label{tab:data}
\end{planotable}
}

\end{document}